\documentclass[prd,nofootinbib,floats,aps,superscriptaddress,preprint, letterpaper]{revtex4}

\usepackage[hypertex]{hyperref}
\usepackage{graphicx,array} 
\usepackage{amsbsy}   
\usepackage{multirow}

\usepackage[disable]{todonotes}

\newcolumntype{S}{>{\centering\arraybackslash} m{.52\linewidth} }
\def\TeV{\, {\rm TeV}}
\def\GeV{\, {\rm GeV}}
\def\eg{\emph{e.g.}}
\def\ie{\emph{i.e.}}

\newcommand{\be}{\begin{eqnarray}}
\newcommand{\ee}{\end{eqnarray}}
\newcommand{\eq}[1]{Eq.~\ref{eq:#1}}
\newcommand {\unit} [1] {\; \mathrm {#1}}
\newcommand {\cL} {\mathcal{L}}
\newcommand{\met}{~/\!\!\!\! E_T}
\newcommand{\mHt}{~/\!\!\!\! H_T}

\def\mstl{m_{\tilde t_{1}}}
\def\msth{m_{\tilde t_{2}}}

\def\tr{{\rm Tr}}

\def\mass2{mass${}^2$}

\def\gev{\, {\rm GeV}}
\def\tev{\, {\rm TeV}}

\def\mass2{mass${}^2$}

\definecolor{nicered}{rgb}{0.7,0.1,0.1}
\definecolor{nicegreen}{rgb}{0.1,0.5,0.1}
\hypersetup{colorlinks,citecolor= nicegreen,linkcolor= nicered}

\begin{document}
\preprint{DESY 11-193}
\preprint{CERN-PH-TH/265}
\pagestyle{plain}

\title{Natural SUSY Endures}

\author{Michele Papucci}
\affiliation{Theoretical Physics Group, 
Lawrence Berkeley National Laboratory, Berkeley, CA 94720}
\affiliation{Department of Physics, University of California,
Berkeley, CA 94720}

\author{Joshua T. Ruderman}
\affiliation{Theoretical Physics Group, 
Lawrence Berkeley National Laboratory, Berkeley, CA 94720}
\affiliation{Department of Physics, University of California,
Berkeley, CA 94720}

\author{Andreas Weiler}
\affiliation{DESY, Notkestrasse 85, D-22607 Hamburg, Germany}
\affiliation{CERN TH-PH Division, Meyrin, Switzerland\\ \mbox{}}

\begin{abstract}

The first 1~fb$^{-1}$ of LHC searches have set impressive limits on new colored particles decaying to missing energy. We address the implication of these searches for naturalness in supersymmetry (SUSY). General bottom-up considerations of natural electroweak symmetry breaking show that higgsinos, stops, and the gluino should not be too far above the weak scale. The rest of the spectrum, including the squarks of the first two generations, can be heavier and beyond the current LHC reach. 
We have used collider simulations to determine the limits that all of the 1~fb$^{-1}$ searches pose on higgsinos, stops, and the gluino.
We find that stops and the left-handed sbottom are starting to be constrained and must be heavier than about 200-300 GeV when decaying to higgsinos. The gluino must be heavier than about 600-800 GeV when it decays to stops and sbottoms. While these findings point toward scenarios with a lighter third generation split from the other squarks, we do find that moderately-tuned regions remain, where the gluino is just above 1 TeV and all the squarks are degenerate and light.  Among all the searches, jets plus missing energy and same-sign dileptons often provide the most powerful probes of natural SUSY. Overall, our results indicate that natural SUSY has survived the first 1~fb$^{-1}$ of data. The LHC is now on the brink of exploring the most interesting region of SUSY parameter space.

\end{abstract}

\maketitle

\tableofcontents

\newpage
\section{Introduction}

The experiments of the Large Hadron Collider (LHC) at CERN are now searching extensively for signals of supersymmetry (SUSY)\@.  So far, the experiments have announced no definitive sign of new physics.  Instead, they have used the first 1~fb$^{-1}$ of data to perform an impressive number of searches that have produced increasingly strong limits on colored superparticles decaying to missing energy~\cite{Aad:2011ib,Aad:2011qa, ATLAS-CONF-2011-098, ATLAS-CONF-2011-130, Collaboration:2011iu, ATLAS-CONF-2011-126,ATLAS-CONF-2011-123,Collaboration:2011wc,ATLAS-CONF-2011-144,Collaboration:2011cw,Collaboration:2011zy,CMS-PAS-SUS-11-004,CMS-PAS-SUS-11-005,CMS-PAS-SUS-11-006,CMS-PAS-EXO-11-036,CMS-PAS-EXO-11-050,CMS-PAS-SUS-11-015,CMS-PAS-SUS-11-010,CMS-PAS-SUS-11-011,CMS-PAS-SUS-11-017,CMS-PAS-SUS-11-013,CMS-PAS-EXO-11-045,CMS-PAS-EXO-11-051}.  These limits have led some to conclude, perhaps prematurely, that SUSY is ``ruled out" below 1 TeV\@.  We would like to revisit this statement and understand whether or not SUSY remains a compelling paradigm for new physics at the weak scale.  If SUSY is indeed still interesting, it is natural to ask: what are the best channels to search for it from now on?  After all, the first fb$^{-1}$ at 7 TeV  were the ``early days'' for the LHC, with many superparticles still out of reach.

We believe that naturalness provides a useful criterion to address the status of SUSY.  Supersymmetry at the electroweak scale is motivated by solving the gauge hierarchy problem and natural electroweak symmetry breaking is the leading motivation for why we might expect to discover superpartners at the LHC.  The naturalness requirement is elegantly summarized by the following tree-level relation in the Minimal Supersymmetric Standard Model (MSSM),
\be \label{eq:tune}
-\frac{m_Z^2}{2} = |\mu|^2 + m_{H_u}^2.
\ee
If the superpartners are too heavy, the contributions to the right-hand side must be tuned against each other to achieve electroweak symmetry breaking at the observed energy scale\footnote{We note that equation~\ref{eq:tune} applies to the tree-level MSSM at moderate to large $\tan \beta$, but, as we will discuss below, similar relations hold more generally.}.  

Eq.~\ref{eq:tune} also provides  guidance towards understanding which superparticles are required to be light, \ie, it defines the minimal spectrum for ``Natural SUSY''. As we review in detail in Sect.~\ref{sec:NaturalPrimer}, the masses of the superpartners with the closest ties to the Higgs must not be too far above the weak scale.
In particular, the higgsinos should not be too heavy because their mass is controlled by $\mu$.  The stop and gluino masses, correcting $m_{H_u}^2$ at one and two-loop order, respectively, also cannot be too heavy.  The masses of the rest of the superpartners, including the squarks of the first two generations, are not important for naturalness and can be much heavier than the present LHC reach.

Naturalness in SUSY~\cite{Barbieri:1987fn, naturalness2,Kitano:2006gv, unnatural,sparticlelimits,focuspoint,strumiaLHC} has  been under siege for quite some time.  The LEP-2 limit on the Higgs mass, $m_h > 114.4\unit{GeV}$~\cite{Barate:2003sz} has led to the so called ``LEP Paradox''/``Little Hierarchy Problem''~\cite{Barbieri:2000gf}.

Here we would like to make a clear distinction between two different types of possible fine-tuning problems that one can consider today:
\begin{enumerate}
\item {\bf Little Hierarchy Problem}: in order to raise the Higgs mass above the LEP-2 limit with radiative corrections, large stop masses are required, $m_{\tilde t} \gtrsim 300 - 1000\unit{TeV}$.  The large stop masses feed into $m_{H_u}^2$ in equation~\ref{eq:tune}, leading to fine-tuning.
\item {\bf Direct LHC Limits}: the stops and gluino masses are directly constrained, leading to fine-tuning in equation~\ref{eq:tune}.
\end{enumerate}

These two fine-tuning problems are intrinsically different, the first being an indirect argument, tightly bound to the MSSM. In fact the model-dependence of the little hierarchy problem is clear when one moves away from the MSSM, as it has been shown in the recent years. For example, the addition of a gauge singlet, as in the NMSSM (see~\cite{Ellwanger:2009dp} and the references therein), can contribute to the Higgs quartic coupling and raise the Higgs mass without introducing fine-tuning~\cite{raisehiggs, Barbieri:2006bg}.  On the other hand new physics can modify the Higgs boson decays in ways that weaken the LEP-2 limit (see the references within~\cite{Chang:2008cw} and more recently~\cite{morehidehiggs}).

On the contrary, the LHC has the potential to probe fine-tuning in a model-independent way by directly placing limits on the superpartner masses. 
The LHC experiments have already presented strong limits on the squarks of the first two generations, constraining them to be heavier than $m_{\tilde q} \gtrsim 700-1000\unit{GeV}$.  If SUSY breaking is mediated to the squarks in a flavor-blind fashion, this limit now drives the fine-tuning in the SSM.  Therefore, naturalness points towards the possibility that the squark soft masses are not flavor degenerate.  Instead, the stops may be significantly lighter than the squarks of the first two generations, if the SUSY breaking mechanism is intertwined with flavor.  This possibility has received serious attention from the theory community for some time~\cite{decouple1st2nd}, and is now also hinted at by the null results of the LHC. 

The above considerations suggest that the most important question about SUSY is to determine the limits on the higgsino, stop, and gluino masses.  However most experimental presentations of SUSY results entangle the limits on these superpartners with the limits on superpartners whose masses do not matter for naturalness, like the squarks of the first two generations. Moreover it is not clear that the present searches that are specifically tailored to gluino and third generation squarks provide the most effective way to search for these states. 

Therefore, in order to ascertain the status of naturalness,  we have used collider simulations to determine the limits on higgsinos, stops, and the gluino. We implemented all the available SUSY searches~\cite{Aad:2011ib, Aad:2011qa, ATLAS-CONF-2011-098, ATLAS-CONF-2011-130, Collaboration:2011iu, Collaboration:2011zy, CMS-PAS-SUS-11-004, CMS-PAS-SUS-11-005, CMS-PAS-SUS-11-015, CMS-PAS-SUS-11-010, CMS-PAS-SUS-11-011, CMS-PAS-SUS-11-017} (and some of the relevant exotica ones~\cite{ATLAS-CONF-2011-126, ATLAS-CONF-2011-123, Collaboration:2011wc, ATLAS-CONF-2011-144, CMS-PAS-EXO-11-036, CMS-PAS-EXO-11-050}) based on approximately $1\unit{fb}^{-1}$.  We cross-checked our results using two different approaches: (1) the fast simulation package PGS~\cite{pgs}, which includes a crude detector simulation with smearing, and (2) our own new pipeline, tentatively called ATOM~\cite{atom}, which uses truth level objects and corrects for  efficiencies of leptons, photons, and b-jets.  The two pipelines are validated against the experimental results of all the analyses that we consider, and their results agree with each other. By using the event yields presented in the experimental papers, we can derive ``theorist's limits'' on natural SUSY, \ie, estimates of what could be excluded by full experimental studies.  Our results provide the benefit of showing which, among the current searches,
sets the strongest limits and what are the weaknesses and strengths of the existing analyses. Such information could be used as a starting point for future experimental investigations. 

In this work, as we will see, we find that the LHC now has the reach to begin to probe the direct production of stops, in certain cases.  There is also the reach to probe the left-handed sbottom, who also must be light because of the Standard Model (SM) weak isospin symmetry.  The reach for gluinos decaying to stops and sbottoms is clearly larger, given its larger production cross-section.  While, {\it a priori}, the gluino mass is less constrained by naturalness than the stop mass because it only contributes to the Higgs potential at two loops, we will see that the limits on the gluino are now comparably important, for naturalness, as the limits on stops, given the larger gluino cross-section. At the same time we find no reach yet to directly probe the higgsino mass beyond the LEP-2 limit on charginos~\cite{LEPsusyWork}.

A number of studies have already considered the implications of the LHC results for SUSY.  On one side there have been studies interpreting the results in terms of specific UV models, such as CMSSM/mSUGRA with 35~pb$^{-1}$~\cite{Msugra35pb} and 1~fb$^{-1}$\cite{Msugra1fb}, and anomaly mediation with 1~fb$^{-1}$\cite{amsb1fb},  focusing on characteristic theory-based slices of the soft breaking parameter space. On the other side, there have been bottom-up studies based on broad parameter scans with 35~pb$^{-1}$~\cite{scan35pb} and 1~fb$^{-1}$~\cite{scan1fb}, trying to cover the whole MSSM parameter space systematically, agnostic of any theoretical bias.  

\begin{figure}[h!]
\label{fig:NaturalSpectrum}
\begin{center} \includegraphics[width=0.9\textwidth]{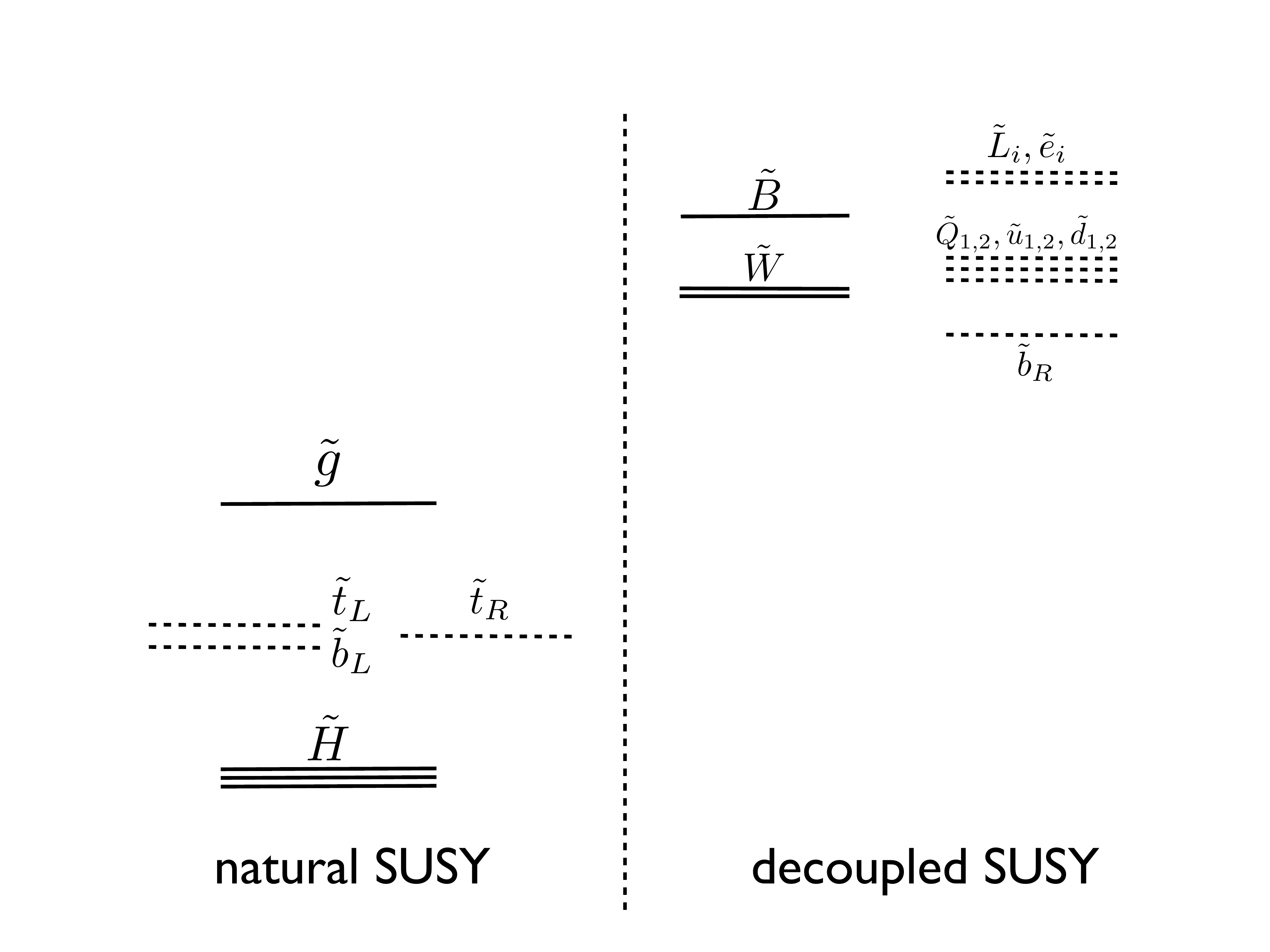} \end{center}
\caption{Natural electroweak symmetry breaking constrains the superpartners on the left to be light.  Meanwhile, the superpartners on the right can be heavy, $M \gg 1$~TeV, without spoiling naturalness.  In this paper, we focus on determining how the LHC data constrains the masses of the superpartners on the left.}
\end{figure}

Our approach differs in that it is decidedly bottom-up, but more focused than broad scans which are penalized by the ``curse of dimensionality'' of the SUSY parameter space. We determine the limits on superpartner masses specified in terms of soft parameters at the electroweak scale.  We restrict the dimensionality of the parameter space by adopting a simplified model philosophy~\cite{Alwall:2008ag}, which is to decouple the states that are not relevant for the signature of interest.  Our choices of simplified models are carefully motivated by naturalness because the states that we keep light are required by fine-tuning to be light and the states that we decouple are unrelated to naturalness~\cite{sparticlelimits}, as summarized in Fig.~\ref{fig:NaturalSpectrum}.

The structure of the paper is as follows. In Section 2, we review in more detail the implications of natural electroweak symmetry breaking in SUSY and we derive the implications for the sparticle spectrum. We remark on the little hierarchy problem and the growing preference for flavor dependent supersymmetry breaking. In Section 3, we review the current status of supersymmetry searches, focusing on the results relevant for our discussion on naturalness. Section 4 contains the main results of our paper: our estimated limits on the masses of stops and gluinos. In Section 5, we interpret our results in the context of specific models, such as the MSSM, scenarios with gaugino unification or those with Minimal Flavor Violation (MFV). We conclude in Section 6, briefly summarizing our findings. The appendices contain a detailed description of ATOM and our validation procedure, and a brief discussion about the challenge of estimating the future reach of the searches we have considered.

\section{SUSY Naturalness Primer}
\label{sec:NaturalPrimer}
In this section we review the basic arguments that determine the minimal set of requirements for natural ElectroWeak Symmetry Breaking (EWSB) in a supersymmetric theory. The subject has received a lot of attention in the past decades~
\cite{Barbieri:1987fn,naturalness2,Kitano:2006gv}. Here we will recollect the main points, necessary for the discussions of the following sections. In doing so, we will try to keep the discussion as general as possible, without committing to the specific Higgs potential of the MSSM.  We do specialize the discussion to 4D theories because some aspects of fine tuning can be modified in higher dimensional setups.

In a natural theory of EWSB the various contributions to the quadratic terms of the Higgs potential should be comparable in size and of the order of the electroweak scale $v\sim 246{\GeV}$. The relevant terms are actually those determining the curvature of the potential in the direction of the Higgs vacuum expectation value.
Therefore the discussion of naturalness can be reduced to a one-dimensional problem as in the Standard Model,
\begin{equation} \label{eq:ThePotential}
V= m_{H}^{2}| H |^{2}+\lambda | H|^{4}
\end{equation}
where $m_{H}^{2}$ will be in general a linear combination of the various masses of the Higgs fields with coefficients that depend on mixing angles, {\it e.g.} $\beta$ in the MSSM\@.\footnote{It is straightforward to extend this discussion to include SM singlets that receive vevs, see for example~\cite{Barbieri:2006bg}.} Each contribution, $\delta m_{H}^{2}$, to the Higgs mass should be less than or  of the order of $m_{H}^{2}$, otherwise various contributions need to be finely tuned to cancel each other. Therefore $\delta m_{H}^{2}/m_{H}^{2}$ should not be large. By using $m_{h}^{2}=-2 m_{H}^{2}$ one can define as a measure of fine-tuning~\cite{Kitano:2006gv},
\begin{equation}
\Delta \equiv \frac{2\delta m_{H}^{2}}{m_{h}^{2}}.
\end{equation}
Here, $m_h^2$ reduces to the physical Higgs boson mass in the MSSM in the decoupling regime.  In fully mixed MSSM scenarios, or in more general potentials, $m_h^2$ will be a (model-dependent) linear combination of the physical neutral CP-even Higgs boson masses.  As is well known, increasing the physical Higgs boson mass (\ie ~the quartic coupling) alleviates the fine-tuning~\cite{raisehiggs,Barbieri:2006bg}.

If we specialize to the decoupling limit of the MSSM and approximate the quartic coupling by its tree level value $\lambda \propto (g^{2}+g^{\prime 2}) \cos^2 2 \beta$, then we find that $m_{h}^{2}=  \cos^2 2 \beta \, m_{Z}^{2}$.  We then recover the usual formula for fine tuning in the MSSM, Eq.~\ref{eq:tune}, in the large $\tan \beta$ limit.

In a SUSY theory at tree level, $m_{H}^{2}$ will include the $\mu$ term\footnote{In theories where the $\mu$-term is generated by the vev of some other field, its effective size is generically bound to be of the order of the electroweak scale by naturalness arguments. For a proof in the NMSSM see, \eg,~\cite{Barbieri:2006bg}.}. Given the size of the top quark mass, $m_{H}^{2}$ also includes the soft mass of the Higgs field coupled to the up-type quarks, $m_{H_{u}}$. 
Whether the soft mass for the down-type Higgs, $m_{H_{d}}$, or other soft terms in an extended Higgs sector, should be as light as $\mu$ and $m_{H_{u}}$ is instead a model-dependent question, and a heavier $m_{H_{d}}$ can even lead to improvements~\cite{Dine:1997qj}.
The key observation that is relevant for  SUSY collider phenomenology is that higgsinos must be light because their mass is directly controlled by $\mu$,
\begin{equation}
\mu \lesssim 200\GeV\left(\frac{m_{h}}{120\GeV}\right)\left(\frac{\Delta^{-1}}{20\%}\right)^{-1/2} 
\end{equation}

At loop level there are additional constraints. The Higgs potential in a SUSY theory is corrected by both gauge and Yukawa interactions, the largest contribution coming from the top-stop loop. In extensions of the MSSM there can be additional corrections, e.g. coming from Higgs singlet interactions in the NMSSM, which can be important for large values of the couplings.
The radiative corrections to $m_{H_{u}}^{2}$ proportional to the top Yukawa coupling are given by,
\begin{equation}\label{eq:der1}
\delta m_{H_{u}}^{2}|_{stop}=-\frac{3}{8\pi^{2}}y_{t}^{2}\left(m_{Q_{3}}^{2}+ m_{u_{3}}^{2}+|A_{t}|^{2}\right)\log\left(\frac{\Lambda}{\TeV}\right),
\end{equation}
at one loop in the Leading Logarithmic (LL) approximation (which is sufficient for the current discussion), see e.g.~\cite{Martin:1997ns}. Here $\Lambda$ denotes the scale at which SUSY breaking effects are mediated to the Supersymmetric SM\@. Since the soft parameters $m_{Q_{3}}^{2}$, $m_{u_{3}}^{2}$ and $A_{t}$ control the stop spectrum, as it is well-known, the requirement of a natural Higgs potential sets an upper bound on the stop masses.
In particular one has
\begin{equation}
\label{eq:ft-stop}
\sqrt{ \mstl^{2}+\msth^{2}}\lesssim 600\GeV \frac{\sin\beta}{(1+x_{t}^{2})^{1/2}}\left(\frac{\log\left(\Lambda/\TeV\right)}{3}\right)^{-1/2}\left(\frac{m_{h}}{120\GeV}\right)\left(\frac{\Delta^{-1}}{20\%}\right)^{-1/2} \, ,
\end{equation}
where $x_{t}=A_{t}/\sqrt{\mstl^{2}+\msth^{2}}$. Eq.~\ref{eq:ft-stop} imposes a  bound on the heaviest stop mass. Moreover, for a fixed Higgs boson mass, a hierarchical stop spectrum induced by a large off-diagonal term $A_{t}$ tend to worsen the fine-tuning due to the direct presence of $A_{t}$ in the r.h.s. of eq.~\ref{eq:der1}.

All the other radiative contributions to the Higgs potential from the other SM particles pose much weaker bounds on the supersymmetric spectrum. The only exception is the gluino, which induces a large correction to the top squark masses at 1-loop and therefore feeds  into the Higgs potential at two loops. One finds, in the LL approximation,
\begin{equation}\label{eq:gluino}
\delta m_{H_{u}}^{2}|_{gluino} = -\frac{2}{\pi^{2}}y_{t}^{2}\left(\frac{\alpha_{s}}{\pi}\right)|M_{3}|^{2}\log^{2}\left(\frac{\Lambda}{\TeV}\right)\, ,
\end{equation}
where $M_{3}$ is the gluino mass and we have neglected the mixed $A_{t}M_{3}$ contributions that can be relevant for large A-terms. From the previous equation, the gluino mass is bounded from above by naturalness to satisfy,
\begin{equation}
M_{3} \lesssim 900\GeV \sin\beta \left(\frac{\log\left(\Lambda/\TeV\right)}{3}\right)^{-1}\left(\frac{m_{h}}{120\GeV}\right)\left(\frac{\Delta^{-1}}{20\%}\right)^{-1/2}\, . 
\end{equation}
In the case of Dirac gauginos~\cite{Fox:2002bu} there is only one power of the logarithm\footnote{The other logarithm is traded for a logarithm of the ratio of soft masses. We assume that the new log is ${\cal O}(1)$, but in principle it can be tuned to provide further suppression.}  in Eq.~\ref{eq:gluino}, ameliorating the bound by a factor of $(\log\left(\Lambda/\TeV\right))^{1/2}$ and leading to a bound of roughly $1.4\TeV$ with the above parameters.

For completeness, we give also the upper bounds on the other gauginos:
\begin{equation}
\left(M_{1},M_{2}\right) \lesssim (3\TeV,900\GeV) \left(\frac{\log\left(\Lambda/\TeV\right)}{3}\right)^{-1/2}\left(\frac{m_{h}}{120\GeV}\right)\left(\frac{\Delta^{-1}}{20\%}\right)^{-1/2}\, .  
\end{equation} 
The bino is clearly much less constrained, while the wino is as constrained as the gluino, but only for low-scale mediation models. For the squarks and sleptons there is only a significant bound from the D-term contribution, if $\tr (Y_{i} m_{i}^{2})\neq0$, and it is generically in the $5-10\TeV$ range.

In the MSSM, the upper bound on the stop mass from the requirement of natural EWSB is in tension with the lower bound on the Higgs boson mass, set by the LEP-2 experiments. The physical Higgs boson mass is controlled by the quartic coupling and the relevant radiative corrections are~\cite{delhiggsmass,Haber:1996fp}
\begin{equation}\label{eq:der2}
\delta m_{h}^{2}=\frac{3 G_{F}}{\sqrt 2 \pi^{2}}m_{t}^{4}\left(\log\left(\frac{\overline m_{\tilde t}^{2}}{m_{t}^{2}}\right)+\frac{X_{t}^{2}}{\overline m_{\tilde t}^{2}}\left(1-\frac{X_{t}^{2}}{12\overline m_{\tilde t}^{2}}\right)\right)
\end{equation}
with $\overline m_{\tilde t}$ the average stop mass and $X_{t}=A_{t}-\mu \cot\beta$, where $\mu$ is the supersymmetric Higgs mass parameter. Since at tree level $m_{h}\leq m_{Z}$, requiring $m_{h}\gtrsim 114\GeV$ translates into a lower bound on the average stop mass of about $1.2\TeV$ for $X_{t}\ll \overline m_{\tilde t}$ and about $250\GeV$ for $X_{t}=\sqrt 6 \overline m_{\tilde t}$, where the stop contribution to the Higgs mass is maximized.

Before the start of the LHC this was the strongest, though indirect, lower bound on the stop masses and the main source of  fine-tuning for the MSSM. 
However, this lower bound on the stop masses  does not necessarily apply to generalizations of the MSSM. In fact, as in, \eg, the NMSSM~\cite{Ellwanger:2009dp}, an extended Higgs sector can easily lead to new contributions to the Higgs quartic coupling, raising the Higgs mass above the LEP limit without the necessity of having very heavy stops~\cite{raisehiggs}.

On the other hand, \eq{der1} holds generically, and one can address the question of the naturalness of the electroweak scale in light of direct sparticle searches, independently of the searches for the Higgs boson(s)\footnote{An extended structure of the Higgs sector will also modify the spectrum of the neutralinos and charginos, and change their relative branching ratios into gauge bosons vs. Higgses. These effects can modify, in general, the phenomenology of SUSY searches. However the modifications caused by an extended Higgs sector are most important for searches looking at direct electroweak-ino production, which is beyond the LHC capabilities with $1fb^{-1}$. We therefore neglect this issue in the rest of the paper.}.

Let us now summarize the minimal requirements for a natural SUSY spectrum:
\begin{itemize}
\item two stops and one (left-handed) sbottom, both below $500-700~\GeV$.
\item two higgsinos, \ie, one chargino and two neutralinos below $200-350~\GeV$. In the absence of other chargino/neutralinos, their spectrum is quasi-degenerate.
\item a not too heavy gluino, below $900~\GeV-1.5~\TeV$.
\end{itemize}
There are some model-dependent motivations for augmenting this minimal spectrum with additional light states.  For example, there could also be a light gravitino at the bottom of the spectrum because a low mediation scale is motivated by reducing the size of the logarithm in Eqs.~\ref{eq:ft-stop} and~\ref{eq:gluino}.  Or, there could be an extra light neutralino (such as a bino or singlino) motivated by dark matter.
The rest of the superparticles may all be decoupled. 

The relevant task is to determine the lower bounds on the masses of third generation squarks, the gluino, and higgsinos, coming from direct collider searches, such as the searches that have been performed so far at the $7\TeV$ LHC.  This will be the subject of the following sections.

As we will summarize in the next section, the LHC presently sets the strongest bounds on the production of gluinos and the squarks of the first two generations. Therefore it is worth discussing scenarios where the spectrum of the third generation squarks is lighter than that of the first two generations~\cite{sparticlelimits,decouple1st2nd}.
Scenarios of this type have less tension with naturalness only if the squark masses are introduced in a flavor non-universal way at the scale where SUSY breaking is mediated to the SSM sector. In fact, squark mass splittings induced by renormalization group evolution originate from the same top Yukawa interactions that correct the Higgs potential. 
Therefore, in flavor-blind SUSY mediation models, large splittings between squarks in the IR actually \emph {increases} the fine-tuning in the Higgs potential. In particular, at one loop one has,
\begin{equation}
\delta m_{H}^{2} \simeq 3 \left(m_{Q_{3}}^{2}-m_{Q_{1,2}}^{2}\right) \simeq \frac{3}{2} \left(m_{U_{3}}^{2}-m_{U_{1,2}}^{2}\right) ,
\end{equation}
where the squark mass splittings pose a lower bound on the amount of fine-tuning.
The implications of the LHC results on this class of models will be further discussed in Section~\ref{sec:interpretation}.


\section{Current status of SUSY searches}
\label{sec:status}



In this section we will study the consequences of the first one and a half years of LHC results on supersymmetry.
The most relevant analyses performed by the ATLAS and CMS collaborations are listed in Table~\ref{tab:searches11}, based on approximately $1fb^{-1}$ of luminosity from the 2011 dataset. The list contains mostly searches for SUSY, but also some exotica searches that were not used to set limits on SUSY, highlighted in blue. Some of the analyses have not been included in this work because they appeared while this work was being completed, and are highlighted in red.

\begin{table}[h!]
\begin{center}
\begin{tabular}{|c||c|c|c||c|c|c|}
\hline
& \multicolumn{3}{|c||}{ATLAS} & \multicolumn{3}{|c|}{CMS} \\
\cline{2-7} \cline{2-7} 
								& channel				&$\cL~[\mathrm{fb}^{-1}]$	& ref. 						& channel 				&$\cL~[\mathrm{fb}^{-1}]$		& ref. \\
\hline		
\multirow{2}{*}{jets~$+ \met$}       		& 2-4 jets 				& 	1.04				&\cite{Aad:2011ib}		                  & $\alpha_T$    			& 1.14 					& \cite{Collaboration:2011zy} \\
 								& 6-8 jets        			&       1.34             		&\cite{Aad:2011qa} 	 			& $H_T, \mHt$				& 1.1						& \cite{CMS-PAS-SUS-11-004} \\	
\hline
\multirow{4}{*}{$b$-jets (+ l's +$\met$)} 	& $1b,2b$				& 	0.83				&\cite{ATLAS-CONF-2011-098}	&$m_{T2}\,\, (+\,b)$ 				& 1.1 					& \cite{CMS-PAS-SUS-11-005} \\
								& $b+1l$				& 	1.03				&\cite{ATLAS-CONF-2011-130}	&\textcolor{red}{$1b,2b$}						& 1.1					& \cite{CMS-PAS-SUS-11-006}\\
								&					& 					&							&\textcolor{blue}{$b' b' \rightarrow b+ l^\pm l^\pm, 3l$}& 1.14				& \cite{CMS-PAS-EXO-11-036}	\\
								&					& 					&							&\textcolor{blue}{$t' t' \rightarrow 2b+l^+ l^-$}		& 1.14				& \cite{CMS-PAS-EXO-11-050}	   \\

\hline
\multirow{6}{*}{multilepton (+$\met$) } 	& $1l$				& 	1.04				&\cite{Collaboration:2011iu}		& 	$1l$					& 1.1	 					& \cite{CMS-PAS-SUS-11-015}  \\
								& \textcolor{blue}{$\mu^\pm \mu ^\pm$}	& 	1.6				& \cite{ATLAS-CONF-2011-126}	&SS dilepton				& 0.98	 				& \cite{CMS-PAS-SUS-11-010} \\
								& \textcolor{blue}{$t \bar t \rightarrow 2l$ }  &	1.04				& \cite{ATLAS-CONF-2011-123}	& OS dilepton        			& 0.98	 				& \cite{CMS-PAS-SUS-11-011} \\
								& \textcolor{blue}{$t \bar t \rightarrow 1l$ }  & 	1.04				& \cite{Collaboration:2011wc}		&$Z\rightarrow l^+ l^-$    		& 0.98					& \cite{CMS-PAS-SUS-11-017} \\
     								& \textcolor{blue}{$4l$}   				&	1.02				& \cite{ATLAS-CONF-2011-144} 	&\textcolor{red}{$3l,4l + \met$}				& 2.1						& \cite{CMS-PAS-SUS-11-013} \\
     								& \textcolor{red}{$2l$} 				& 1.04				& \cite{Collaboration:2011cw}	 	&\textcolor{red}{$3l,4l$}					& 2.1						& \cite{CMS-PAS-EXO-11-045} \\
\hline
\end{tabular} 
\end{center}
\caption{ \label{tab:searches11}
Searches by ATLAS and CMS, with about 1 fb$^{-1}$, for signatures that are produced by models of natural supersymmetry.  We have categorized the searches into three categories, (1) fully hadronic, (2) heavy flavor, with or without leptons, and (3) multileptons without heavy flavor.  The searches with blue labels have not been used by experimentalists to set limits on supersymmetry, but we have included them because they overlap with SUSY signature space.  We have simulated all of the above searches and included them in our analysis, with the exception of the searches with red labels, which were released while we were finalizing this study.  We explored the possibility of using the CMS search for $t'$ in the lepton plus jets channel~\cite{CMS-PAS-EXO-11-051}, however this search uses a kinematic fit on signal plus background and does not report enough information for us to extrapolate this fit to other signals.}
\end{table}

Let us first summarize the results presented by the two collaborations in their papers. This will set the stage for the more general investigation of the natural SUSY parameter space described in the previous section, which will be performed in Section~\ref{sec:limits}.

The performance of nearly all of the SUSY analyses are compared within the standard CMSSM $m_{0}-m_{1/2}$ plane. Here, the most stringent constraints come from the jet +$\met$ searches, and provide limits of $m_{1/2}\gtrsim540~\gev$ for low $m_{0}$ and $m_{1/2}\gtrsim300~\gev$ for large $m_{0}$, corresponding to squark masses of $\sim1.1~\tev$. The analyses requiring one or more leptons, although looking at different final states, tend to provide weaker constraints in this plane.

However, for this study it is much more instructive to extract information from the simplified model presentation of the above analyses. For the case of squarks and gluinos, ATLAS presents the results in a squark-gluino-neutralino simplified model~\cite{Aad:2011ib}, with two free parameters, $m_{\tilde q}=M_{Q_{1,2}}=M_{D_{1,2}}=M_{U_{1,2}}$ and $m_{\tilde g}$, with $m_{\chi^{0}}=0$, thus maximizing the multiplicity of squarks and loosing the dependence of the bounds on the neutralino mass. CMS instead presents two separate plots, one for squark pair production, with each squark decaying into a quark and a neutralino, the other for gluino pair production, with each gluino three-body decaying into two quarks and a neutralino, using $(m_{\tilde q},m_{\chi^{0}})$ and  $(m_{\tilde g},m_{\chi^{0}})$ as parameters, with all the other states decoupled.
This can allow the exploration of more general squark spectra and shows the dependence on the neutralino mass, but at the same time misses the associated squark-gluino production relevant when $m_{\tilde q}\sim m_{\tilde g}$, which is instead captured by the ATLAS presentation. Nevertheless, in both cases one can easily extrapolate the available information.
One finds that squarks and gluinos decaying hadronically are constrained to be at or above $900~\gev-1~\tev$, imposing strong constraints on flavor universal models, as explained in the previous section. There are however ways out of this result, as can be seen from the CMS simplified model summary plot~\cite{CMSsimpified}, which presents the dependence of the CMS limits on the Lightest Supersymmetric Particle (LSP) mass: the bounds get obviously weaker when the separation between the squark (gluino) and the neutralino is compressed, because events become less energetic.
In particular, for the case of squarks decaying into a jet and a neutralino, the CMS $\alpha_{T}$ search~\cite{Collaboration:2011zy} sets a lower limit of $500\gev$ on $m_{\tilde q}$ for $m_{\tilde q}-m_{\chi}=200\gev$ and decoupled gluino. Clearly this is an important point: a quasi-degenerate squark spectrum around $500\gev$ with $\mu=300\gev$ is only moderately tuned, and does not necessitate the introduction of any splitting between the first/second and the third generation squarks. The question here is how heavy must the gluino be for this result to hold, and whether or not other searches impose stronger constraints  on the squark masses. We will address this issue in Sect.~\ref{sec:interpretation}.

Let us now move to briefly discuss the searches requiring b-jets. These includes both SUSY and exotica ($t'$) searches. Different analyses require different numbers of leptons in the final state and/or the presence of $\met$. In particular, the CMS $M_{T_{2}}$ analysis~\cite{CMS-PAS-SUS-11-005} is a looser jets+$\met$ search where the cuts on the hadronic activity $H_{T}$ and the $\met$ have been relaxed in favor of the requirement of a b-jet. ATLAS, on the other hand, has presented two analyses, tailored at gluino pair production with gluinos decaying either to sbottoms~\cite{ATLAS-CONF-2011-098} or to stops~\cite{ATLAS-CONF-2011-130}, requiring 0 or 1 leptons, respectively. They also present their results in terms of simplified models, parameterized by gluino and sbottom (stop) masses or, in case of~\cite{ATLAS-CONF-2011-130}, gluino/neutralino masses for a simplified model where the gluino decays three-body, $\tilde g \rightarrow t \bar t \chi^{0}$. One can see that their limits are driven by gluino pair production and that they disappear for a sufficiently heavy gluino, $m_{\tilde g}\gtrsim 500-600\gev$. On the other hand it is not clear whether other searches of Table~\ref{tab:searches11} also have the power to constrain these scenarios and what is their reach. This is the main motivation for the study of the next section, where we will consider the constraints on stops and stops+gluino, decaying into higgsinos (and/or a bino).

Finally various multi-lepton searches with and without missing energy have the power to constrain  scenarios involving decays into tops and gauge bosons, since these states may yield leptons in the final state.  With leptons, the SM backgrounds are considerably smaller than those for the jets+MET searches. Therefore it is interesting to see whether these analyses are relevant for constraining natural SUSY spectra. Unfortunately this information cannot easily be extracted from the experimental papers, where most of the results are expressed as CMSSM exclusion regions or in simplified models involving first two generation squarks (gluinos) decaying into charginos and neutralinos. Therefore we will investigate the reach of these searches for natural SUSY spectra involving third generation squarks in the next section.

An important set of searches relevant for the limits on third generation squarks are those looking for Heavy Stable (or long-lived) Charged/Colored Particles (HSCPs). Both ATLAS~\cite{Aad:2011yf,Aad:2011hz} and CMS~\cite{CMS-PAS-EXO-11-020, CMS-PAS-EXO-11-022} have performed searches for HSCPs, and the current most stringent limits are around $600\gev$ for stop LSP, which already constitutes moderate fine-tuning. Therefore in the next section we will not consider the possibility of a long-lived stop/sbottom at the bottom of the SUSY spectrum.  Instead, we will always assume that either a higgsino, or in certain cases a bino (gravitino), is the LSP.

Finally, let us comment on the current bounds on direct higgsino production. The most robust limit comes from the LEP-2 constraint on chargino pair production and is about $100~\gev$~\cite{LEPsusyWork}. Collider searches from the Tevatron do constrain charginos and neutralinos from trilepton studies~\cite{Forrest:2009gm}, but the Tevatron only improves on the LEP limit when there are light sleptons in the spectrum,  increasing the number of leptons in the final state.    
On the other hand, the LHC, with less than or about $1fb^{-1}$, probably does not have enough luminosity to produce competitive constraints on the direct production of charginos/neutralinos. Since the Tevatron bounds are model-dependent and sleptons are not required to be light by naturalness,  we will only consider, in the next section,  higgsinos in association with stops (sbottom) with and without the gluino, and use the LEP-2 limit as a lower bound for the higgsino masses.

\section{The Limits}
\label{sec:limits}

In this section we present our main results: our estimates of the limits on the masses of the superpartners that must be light for natural electroweak symmetry breaking.  In order to avoid excessive fine-tuning, the higgsinos, stops, and gluino must not be too heavy, as discussed in Sect.~\ref{sec:NaturalPrimer}.  
We find no LHC limit, with the first $1\unit{fb}^{-1}$, on higgsinos, beyond the LEP-2 limit on the charginos, $m_{\tilde H^\pm} \gtrsim 100$~GeV~\cite{LEPsusyWork}. We do find that the LHC sets limits on the direct production of third generation squarks, and on the production of gluinos that decay through on or off-shell stops and sbottoms.
After briefly discussing our methodology, we present our estimates of the limits on stops in Sect.~\ref{subsec:stop} and on gluinos in Sect.~\ref{subsec:gluino}.

The LHC experiments have not yet presented limits on the direct production of stops or sbottoms decaying to a neutralino LSP.  And only a handful of searches, looking for b-jets plus missing energy, have presented limits on gluinos decaying through on or off-shell stops and sbottoms~\cite{ATLAS-CONF-2011-098,ATLAS-CONF-2011-130}.  However, many searches have been conducted with $1\unit{fb}^{-1}$, as reviewed in Sect.~\ref{sec:status}, and these searches collectively cover a large signature space.  In order to address the status of naturalness in supersymmetry, we would like to ask the question: do the existing LHC searches, conducted with $1\unit{fb}^{-1}$, set limits on the direct production of stops and sbottoms?  And what is the strongest limit on the gluino mass, when stops are light?  In order to answer these two questions, we have simulated the existing searches and estimated the limits on stops and gluinos.   

\subsection{Methodology and Caveats}
\label{subsec:methodology}

Here, we briefly discuss our methodology for simulating the LHC searches.  We calculate the SUSY spectrum and the decay tables for SUSY particles with the program SUSY-HIT~\cite{Djouadi:2006bz}. Events were simulated using Pythia v.~\mbox{6.4.24}~\cite{pythia}.  We use NLO K-factors, from Prospino v.~\mbox{2.1}~\cite{prospino}, for colored superparticles production.  We then pass the events through two different pipelines, allowing us to internally cross-check our results.  The first pipeline, Atom~\cite{atom}, uses truth-level objects and will be further discussed in Appendix~\ref{app:atom}.  As a second pipeline, we use PGS~\cite{pgs}, which acts as a crude detector simulation including smearing.  As we discuss in Appendix~\ref{app:validation}, we validate both simulations by reproducing the published limits of all of the searches.  Typically, both Atom and PGS reproduce experimental acceptances with an accuracy better than 50\%, which results in superparticle mass limits that are normally accurate within about 50 GeV.\footnote{The limit on stops or gluinos is normally not very sensitive to a $\lesssim 50\%$ error in acceptance, because cross-sections are steep functions of masses.  An important exception, to keep in mind, arises when the acceptance is also a steep function of mass, in which case $\sigma \times \epsilon \times \mathrm{A}$ may vary more slowly with mass, enhancing the sensitivity of the limit to mis-modeling the acceptance. One of our pipelines, Atom, automatically detects such cases, allowing us to identify potential problems.}   For searches with multiple channels, we quote the limit from the channel with the best expected limit, at each point in signal parameter space.  All limits are 95\% confidence level exclusions using the $CL_s$ statistic~\cite{CLs}.

In each of the cases considered in the following, we will adopt the simplified model philosophy~\cite{Alwall:2008ag}, which is to only consider the relevant particles (stop/sbottom and higgsino) and to decouple the rest of the spectrum, in order to highlight the relevant kinematics.  We choose 3 TeV as the mass scale for the rest of the decoupled superpartners. Throughout this work, we fix $\tan \beta =10$.

We would like to stress a caveat, inherent to ``theorist level" extrapolations of LHC limits.  It is important to keep in mind that our limits do not represent actual experimental limits.  Accurate limit setting requires the full experimental detector simulation, which we do not have access to, and a careful study of systematic uncertainties of the  signal acceptance, which we do not attempt.  We are not trying to replace these important steps.  Rather, the limits we quote should be viewed as representative estimates of what we believe will be possible to exclude, if the experimentalists apply the current searches to study natural supersymmetric spectra.  We have identified parameter spaces that are useful for assessing the status of naturalness in supersymmetry, and we hope the task of setting more accurate limits on natural supersymmetry will now be taken up by our experimental colleagues.

\subsection{Stop Limits}
\label{subsec:stop}

Stops must be light if electroweak symmetry breaking is natural, because they contribute to $m_{H_u}^2$ at one-loop order.  As we discuss in this section, we have found that the existing LHC searches in certain cases place limits on the direct production of stops.  These limits being as strong as $m_{\tilde t} \gtrsim 300$~GeV, show that the null results of the LHC are starting to  directly probe SUSY naturalness.  Note, that loops of light stops can modify the higgs production cross-section and branching ratios (see~\cite{Djouadi:2005gj} and references therein), and for some choices of parameters, there can be an increase of  $\sigma_{g g \rightarrow h} \times \mathrm{Br}(h \rightarrow \gamma \gamma)$.  This means that LHC Higgs searches can also be used to provide indirect limits on light stops.  We do not consider such indirect limits here, since they rely on model-dependent assumptions about the Higgs sector, and we instead choose to focus on the direct limits on stop production.

Before starting to review the limits, let us recall how the stop masses are determined by soft supersymmetry breaking parameters~\cite{Martin:1997ns}.  In general, left and right-handed stops mix, and the squared stop soft masses are given by the eigenvalues of the following matrix,
\be \label{eq:stopmass}
\left( \begin{array}{ccc}
m_{Q_3}^2 + m_t^2 + t_L m_Z & m_t X_t \\
m_t X_t & m_{U_3}^2 + m_t^2 + t_R m_Z^2
\end{array} \right),
\ee
where $m_{Q_3}$ and $m_{u_3}$ are the left and right-handed stop soft masses, respectively, $X_t = A_t - \mu / \tan \beta$ determines the left-right stop mixing, and $t_{L,R}$ parameterize $D$-term corrections that are introduced by electroweak symmetry breaking.  The $D$-term coefficients are given by $t_L = (1/2-2/3 \sin^2 \theta_W) \cos 2 \beta$ and $t_R = 2/3 \sin^2 \theta_W \cos 2 \beta$.  

As explained before,
naturalness also requires a light left-handed sbottom, whose mass is also determined by $m_{Q_3}$.  If $\tan \beta$ is not too large,  then left-right sbottom mixing can be neglected and the right handed sbottom is not required, by naturalness, to be light.  In this case, the left-handed sbottom mass is given by,
\be \label{eq:sbottommass}
m_{\tilde b_L}^2 = m_{Q_3}^2 + m_b^2 - \left( \frac{1}{2} + \frac{1}{3} \sin ^2 \theta_W \right) \cos 2 \beta \, m_Z^2 ,
\ee
where the last term corresponds to the $D$-term contribution to the sbottom mass.

\begin{figure}[h!]
\label{fig:StopHiggsinoScheme}
\begin{center} \includegraphics[width=0.9 \textwidth]{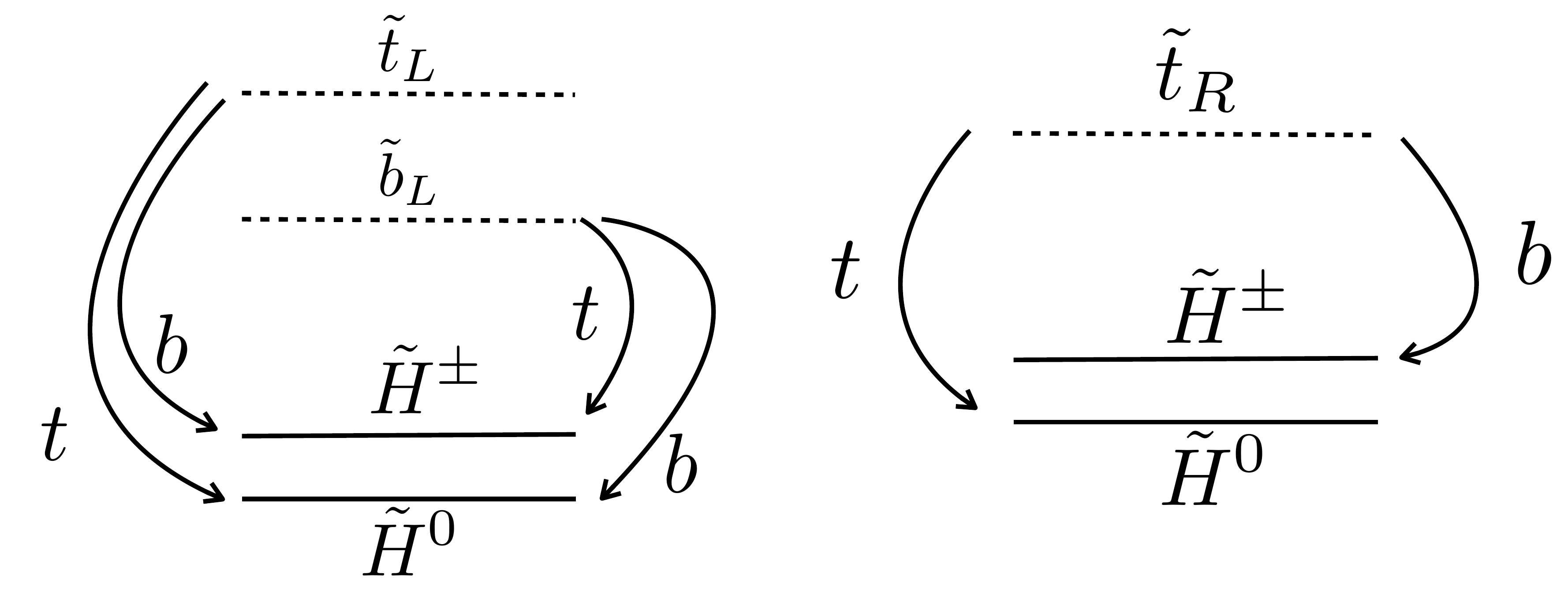} \end{center}
\caption{Possible decay modes in the simplified model consisting only of a left-handed stop/sbottom, or right-handed stop, decaying to a higgsino LSP\@.  On the {\it left}, we show decays of the left-handed stop and left-handed sbottom, whose masses are both determined by $m_{Q_3}$.   On the {\it right}, we show possible decays of the right-handed stop, whose mass is determined by $m_{u_3}$.   At this stage, we neglect left-right stop mixing. }
\end{figure}

We begin by considering the limits on stops, and the left-handed sbottom, with a higgsino LSP.   These are the most important superparticles to be light if supersymmetry is natural.  The spectrum, and the relevant decays, are shown in Fig.~\ref{fig:StopHiggsinoScheme}.   We begin, for simplicity, by neglecting left-right stop mixing, $X_t = 0$ (we will relax this assumption below).  Then, the right-handed stop mass is determined by $m_{u_3}$ and the left-handed stop and sbottom have masses close to $m_{Q_3}$, with the left-handed stop a bit heavier than the left-handed sbottom, due to the $m_t^2$ contribution to the upper-left entry of the stop mass matrix (see eq.~\ref{eq:stopmass}).  As a further simplification, to illustrate the main kinematical features, we separately consider the limits of the left-handed stop/sbottom, and right-handed stop.

\begin{figure}[h!]
\begin{center}  \includegraphics[width=1\textwidth]{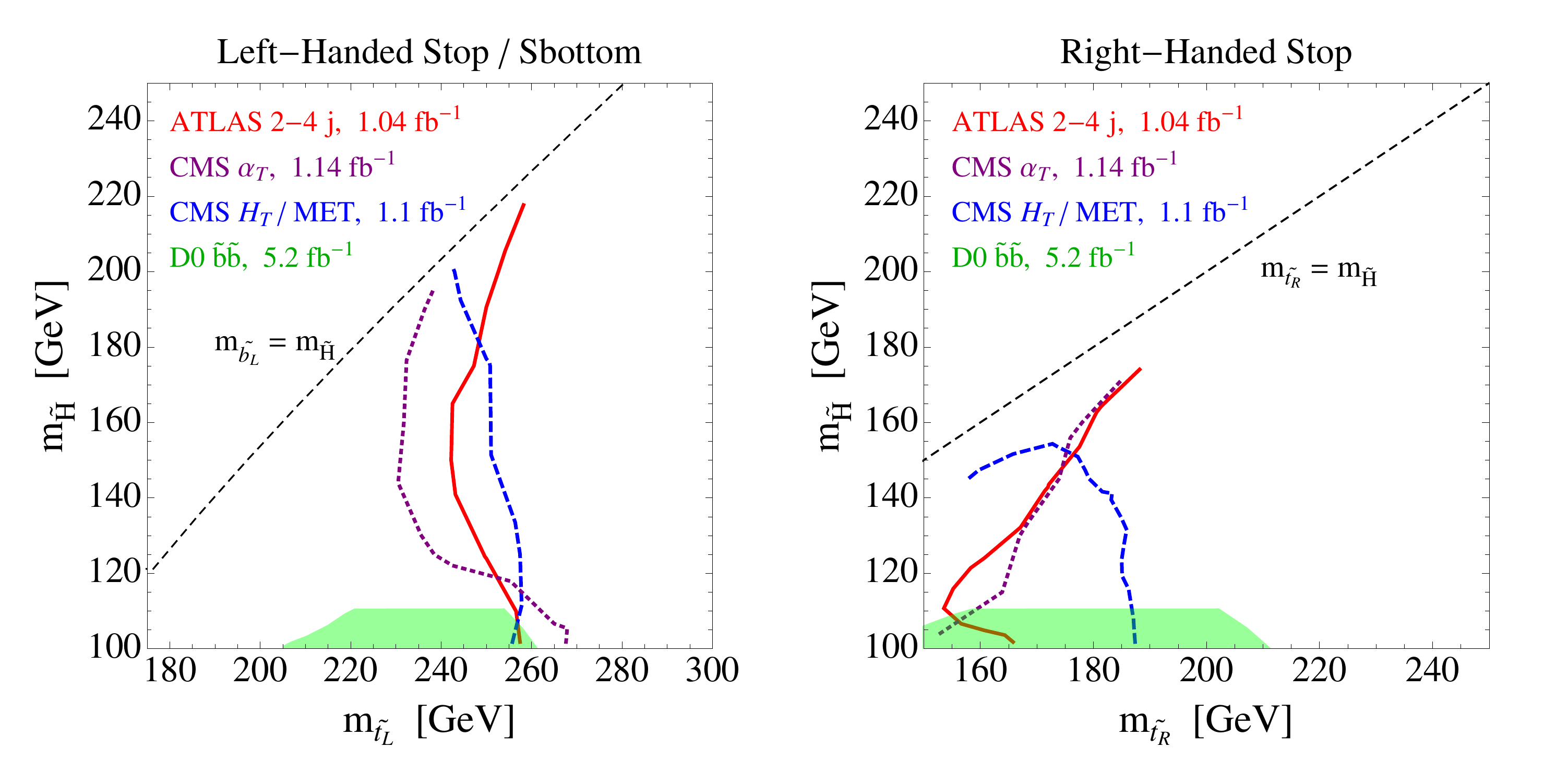}   \end{center}
\caption{ \label{fig:StopHiggsinoLimit}
The LHC limits on the left-handed stop/sbottom ({\it left}) and right-handed stop ({\it right}), with a higgsino LSP\@.  The axes correspond to the stop pole mass and the higgsino mass.   We find that the strongest limits on this scenario come from searches for jets plus missing energy.  For comparison, we show the $D0$ limit with $5.2\unit{fb}^{-1}$ (green), which only applies for $m_{\tilde N_1} \lesssim 110$~GeV, and has been surpassed by the LHC limits.}
\end{figure}

The LHC limit on the left-handed stop and sbottom (right-handed stop) is shown to the left (right) of Fig.~\ref{fig:StopHiggsinoLimit}, respectively.  We find that the strongest limit comes from searches for jets and missing energy, which are shown in the plot.  There is a stronger limit on the left-handed stop than the right-handed stop, because of the additional presence of a sbottom, in the left-handed case, leading to an overall larger production cross-section than for the right-handed stop.  
In both cases the limits are set by both stops and bottoms decaying to b-jets and chargino or neutralino respectively.

We comment that near the edge of the limit, the typical acceptance of the jets plus missing energy searches for this signal is only $\sim \mathcal{O}(10^{-3})$.  This is the right order of magnitude to set a limit because 200 GeV stops have a production cross-section of about 10 pb, which then leads to 10's of events after cuts, in $1\unit{fb}^{-1}$.

To understand why the acceptance is  $\sim \mathcal{O}(10^{-3})$, we consider, as an example, the high missing energy selection of the CMS jets plus missing energy search~\cite{CMS-PAS-SUS-11-004}.  This search demands $H_T >350$~GeV and $\met > 500$~GeV.  We find that moderately hard initial state radiation is required for stops and sbottoms in the mass range of 200-300 GeV to pass this cut.  The low acceptance is related to the probability to produce sufficiently hard radiation.  In order to verify that the acceptance is not considerably underestimated due to the fact that the additional jets are populated only by the parton shower in events generated by Pythia (with the total cross-section normalized to the NLO value), we have also generated events in Madgraph~\cite{madgraph} with stop and sbottom pair production  including also the possibility of radiating one extra parton at the level of the matrix element. Overall we find  good agreement between the two estimates, within our typical uncertainties.

For comparison with the LHC limits, we have also shown in Fig.~\ref{fig:StopHiggsinoLimit}, the strongest limit from the Tevatron, which comes from the 
$D0$ sbottom search with $5.2\unit{fb}^{-1}$.  This search sets limits on sbottom pair production, with the decay $\tilde b \rightarrow b \tilde N_1$.  For the left-handed spectrum, this limit applies directly to the sbottom, which decays $\tilde b_L \rightarrow b \tilde H^0$ for the mass range of interest (the decay to top and chargino is squeezed out).  For the right-handed stop, the dominant decay is $\tilde t_R \rightarrow b \tilde H^\pm$, which means that the stop acts like a sbottom, from the point of view of the Tevatron search\footnote{In order to apply the Tevatron sbottom limit to right-handed stops, we have assumed that the decay products of the charged higgsino are soft enough not to effect the selection, which applies when the mass splitting between the charged and neutral higgsino is small}.  We note that the Tevatron limit  only applies for higgsinos just above the LEP-2 limit, $m_{\tilde H} < 110\unit{GeV}$, and we see that the Tevatron has been surpassed by the LHC in this parameter space.

\begin{figure}[h!]
\begin{center} \includegraphics[width=0.75 \textwidth]{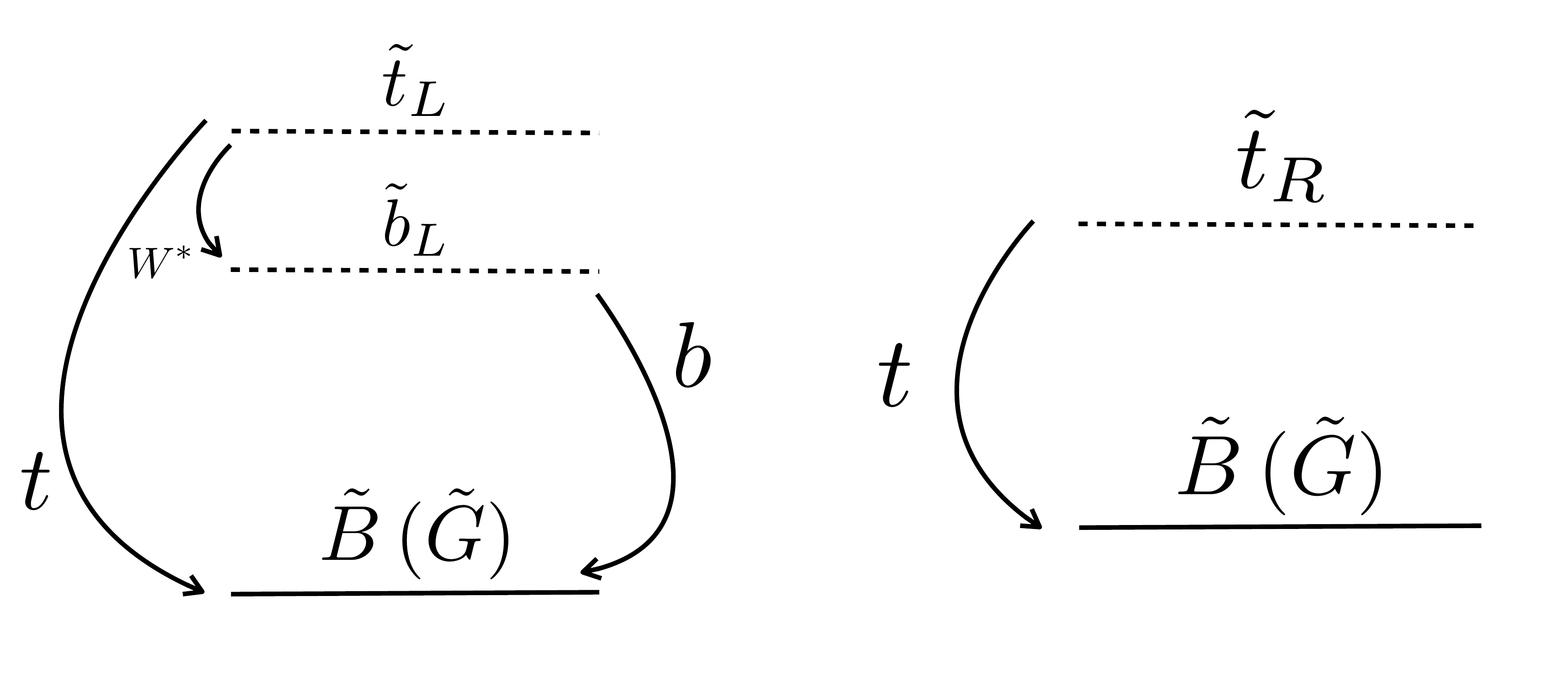} \end{center}
\caption{
\label{fig:StopBinoScheme}
Possible decay modes of the left-handed stop/sbottom ({\it left}), or right-handed stop ({\it right}), to a bino or gravitino LSP.  Higher body final states occur when the mass splittings squeeze out the two-body decays of the stops, $m_{\tilde t_{L,R}} < m_{\tilde B} - m_t$.
}
\end{figure}

We now consider the LHC limit on stops and the left-handed sbottom decaying to a bino (or gravitino) LSP.  Here we will take the higgsinos to be heavier than the stops, and again we neglect left-right stop mixing for simplicity, $X_t=0$.  The relevant spectra and decay modes are shown in Fig.~\ref{fig:StopBinoScheme}.  The most important change, versus higgsino LSP, is that there is no light chargino for the stops and sbottoms to decay to.  For left-handed stops, this means that once the decay to the bino and a top is squeezed out, $m_{\tilde t_L} < m_{\tilde B} + m_t$, the left-handed stop dominantly decays to the sbottom through a 3-body decay, $\tilde t_L \rightarrow W^* \tilde b_L$.  For the right handed stop, once the two body decay is unavailable, $m_{\tilde t_R} < m_{\tilde B} + m_t$, the dominant decay is a three-body decay through an off-shell top.  And once the mass splitting between the stop and the bino is less than the $W$ mass, the dominant decay is 4-body with the top and the $W$ both off-shell.  The right-handed stop decays are challenging to constrain because the final states are similar to the $t \bar t$ background.  The same decay modes apply both for bino and gravitino LSP, the only relevant difference is that the bino mass is a free parameter, whereas the gravitino must be light, $m_{\tilde G} \lesssim \unit{keV}$ for decays to occur within the detector.

\begin{figure}[h!]
\begin{center} \includegraphics[width=0.5 \textwidth]{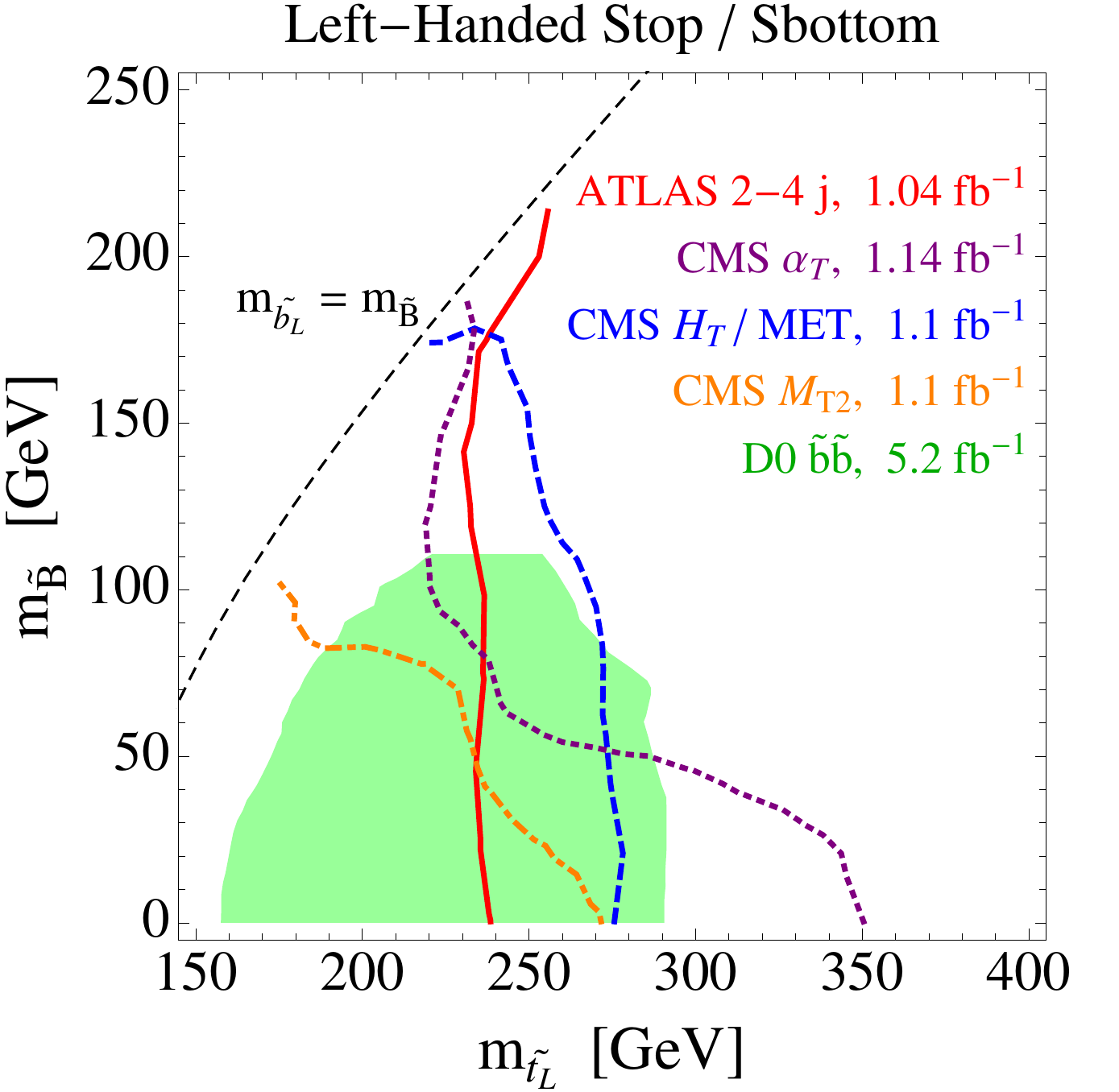} \end{center}
\caption{
\label{fig:StopBinoLimit}
The LHC limits on left-handed stop/sbottom, with a bino LSP\@.  The axes correspond to the stop pole mass and the bino mass.  The limit with a gravitino LSP in place of the bino can be inferred from looking at the line with $m_{\tilde B} \approx 0\unit{GeV}.$  We find that searches for jets plus missing energy set the strongest limits, which surpass the $D0$ limit with $5.2\unit{fb}^{-1}$ (green).  We do not show the case with a right-handed stop with bino/gravitino LSP, where we find no limit above $m_{\tilde t} \gtrsim 200\unit{GeV}$.  We find that there may be marginal sensitivity for lighter right-handed stops, although this requires further investigation due to the similarity of the stop signal and the irreducible top background.}
\end{figure}

We present our estimate of the limit on the left-handed stop/sbottom with bino LSP in Fig.~\ref{fig:StopBinoLimit}.  The limit with a gravitino LSP can be inferred by looking along the $m_{\tilde B} \approx 0$ line of the mass plane.  We find that the strongest limits come from searches for jets plus missing energy, as in the case with the higgsino LSP.  The physics is similar to the higgsino case, except that more of phase space is relevant, since there is no LEP-2 limit on the bino mass.  In the massless bino / gravitino case, we find that the limit on the left-handed stop extends as far as $\sim350$~GeV for light bino, where more phase space is available.

For the right-handed stop decaying to a bino, we show no plot because we find no limit above $m_{\tilde t_R} \gtrsim 200$~GeV.  We do find that there may be marginal sensitivity for stop masses around 200 GeV.  This marginal sensitivity comes from searches for jets plus missing energy, Z plus jets plus $\met$ and from searches for top partners.  We do not show an estimate for the limit on right-handed stops with masses near the top mass because the signal topology is very similar to the $ t \bar t$ background.  This means that any limit extrapolation is sensitive to the detailed systematics of the top background, and we believe this parameter regime requires further dedicated study.

\begin{figure}[h!]
\begin{center} \includegraphics[width=0.8 \textwidth]{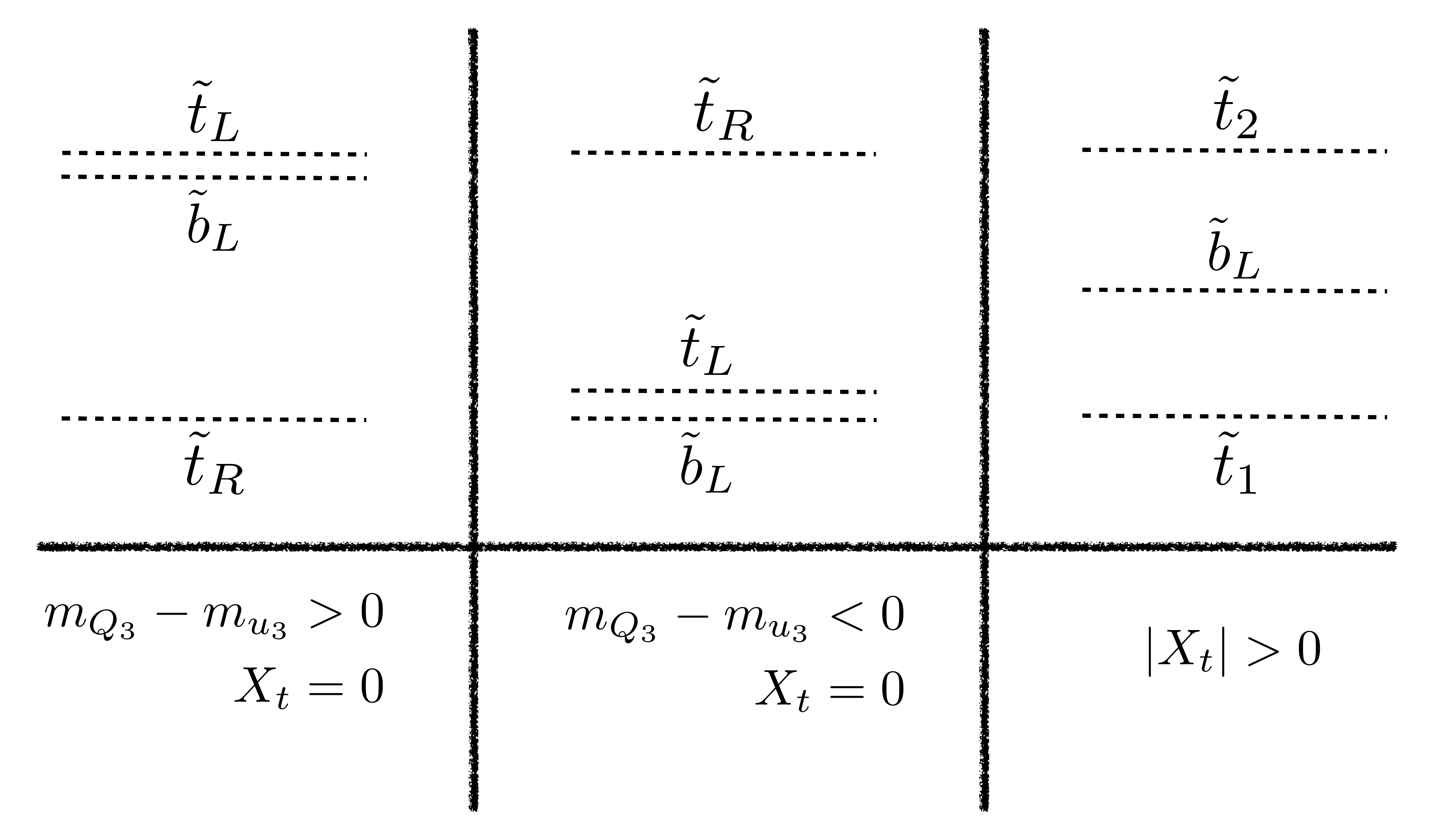} \end{center}
\caption{Different ways that stops can be split and mixed.  The left and right-handed stop pole masses can be split by choosing different soft terms, $m_{Q_3} \neq m_{u_3}$, as shown to the {\it left} and {\it center}.  The stop masses can also be split due to left-right stop mixing, which is controlled by the parameter $X_t$, as shown to the {\it right}.  The left-handed sbottom mass is determined only by $m_{Q_3}$, in the limit that left-right sbottom mixing can be ignored, which we assume here.
\label{fig:SplitStops}
}
\end{figure}

We conclude our discussion of limits on stop production by considering the limit on both left and right-handed stops, including left-right mixing.  By inspecting the stop mass matrix, eq.~\ref{eq:stopmass}, we see that there are two ways to change the relative stop masses, which are depicted in Fig.~\ref{fig:SplitStops}.  The first way is to assign different soft masses for the left and right-handed stops, as shown to the left, and center of Fig.~\ref{fig:SplitStops}.  In the limit of no left-right mixing, the left-handed sbottom and left-handed stop masses will both be close to the value of $m_{Q_3}$ (up to $m_t$ and $D$-term corrections).  The second way to change the stop masses is to introduce left-right stop mixing, $|X_t|>0$, shown to the right of Fig.~\ref{fig:SplitStops}.  When there is large left-right mixing, the sbottom mass is no longer required to be close to one of the stop masses.

\begin{figure}[h!]
\begin{center} \includegraphics[width=1 \textwidth]{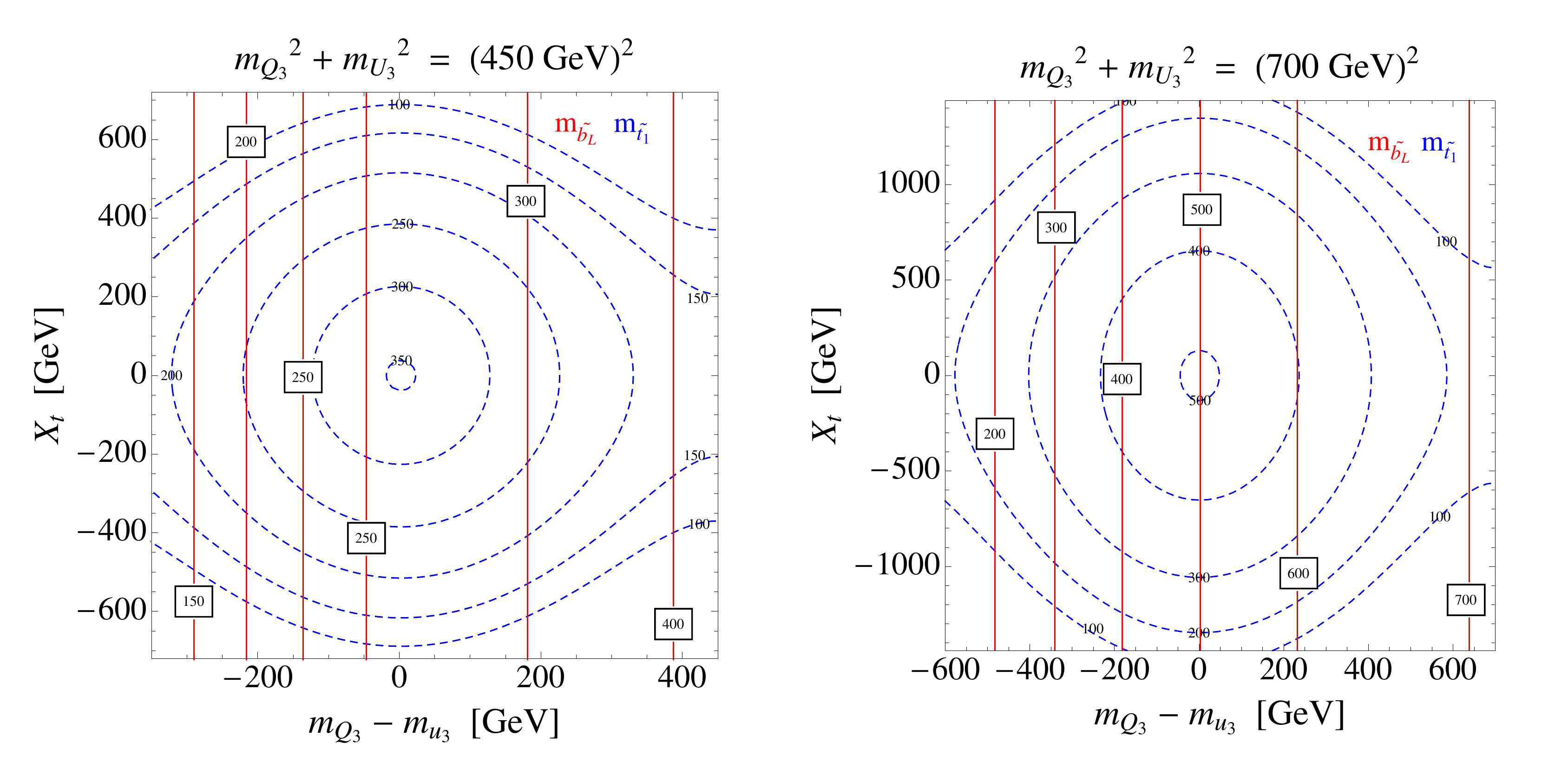} \end{center}
\caption{
\label{fig:StopSbottomMass}
The masses of the lightest stop, $\tilde t_1$, and left-handed sbottom, $\tilde b$, while varying the stop mixing parameter, $X_t$, and the difference of the left and right-handed soft terms, $m_{Q_3} - m_{u_3}$.  Here we take $m_{Q_3}^2+m_{u_3}^2=\left( 450 \unit{GeV} \right)^2$ on the {\it left}, and $\left( 700 \unit{GeV} \right)^2$ on the {\it right}.  Fixing this combination keeps constant the amount of fine-tuning introduced by the stop soft masses.  Moving from left to right, the sbottom mass increases with $m_{Q_3}$.  Meanwhile, the lightest stop mass decreases moving radially outward in the plot, due to different left-right soft masses in the horizontal direction, and left-right mixing in the vertical direction.}
\end{figure}

We have chosen a parameter space designed to study how the LHC limit depends on left-right stop mixing.  We fix the value of $m_{Q_3}^2+m_{u_3}^2 =\left( 450 \unit{GeV} \right)^2$, which fixes the amount of fine-tuning introduced by the stop masses into $\delta m_{H_u}^2$.  Then, we separately vary the difference of the left-right stop soft masses, $m_{Q_3} - m_{u_3}$, and the left-right mixing, $X_t$.  We show how the lightest stop mass and sbottom mass depend on these parameters in Fig.~\ref{fig:StopSbottomMass}.  The sbottom mass increases with $m_{Q_3}$, moving from left to right across the plot.  Meanwhile, the lightest stop mass decreases as either the stop mixing is increased, or as the difference of the stop soft masses is increased.

\begin{figure}[h!]
\begin{center} \includegraphics[width=0.7 \textwidth]{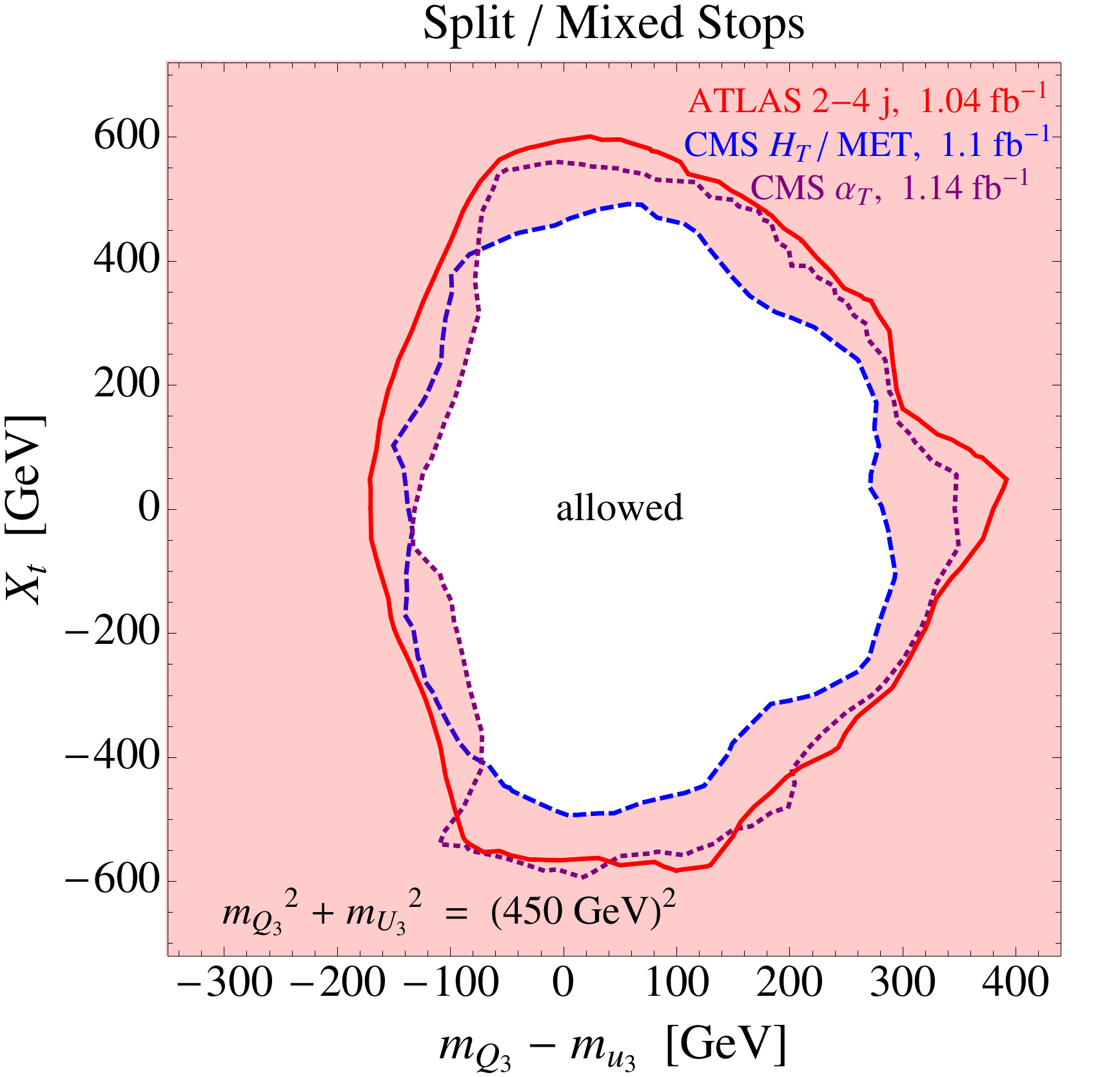} \end{center}
\caption{
\label{fig:DmXtmu}
The limit on the stops and left-handed sbottom, including stop mixing.  We take $m_{Q_3}^2+m_{u_3}^2=\left( 450 \unit{GeV} \right)^2$, which fixes the amount of fine-tuning that the stop soft masses introduce to electroweak symmetry breaking.  We vary the stop mixing, $X_t$, and the difference of the stop soft masses, $m_{Q_3} - m_{u_3}$.  The resulting stop / sbottom mass spectrum is shown in Fig.~\ref{fig:StopSbottomMass}.  The strongest limits come from searches for jets plus missing energy, which exclude the region {\it outside} of the circular exclusion contour.  This is the part of parameter space where one stop becomes light, as shown in Fig.~\ref{fig:StopSbottomMass}.  The green band to the left of the plot is excluded by the $D0$ b-jets plus missing energy search with $5.2\unit{fb}^{-1}$.
}
\end{figure}

We show our LHC limit for this parameter space in Fig.~\ref{fig:DmXtmu}.  Here, we have chosen a higgsino LSP with a mass of 100 GeV.  We note that left-right stop mixing can allow decays between the stops to a Higgs boson, $\tilde t_2 \rightarrow h t_1$.  These decays are clearly more model dependent since we do not have much information on the structure of the Higgs sector yet. For concreteness, we have fixed $m_h = 120$~GeV and take the decoupling limit in the Higgs sector, $m_A \gg m_Z$.  The strongest limit in this parameter space comes again from searches for jets plus missing energy, and the outer parts of the plot are excluded.  This is simple to understand: the exclusion corresponds to the part of parameter space where the lightest stop mass falls below the limit, $m_{\tilde t_1} \gtrsim 200-250$~GeV.  The limits are stronger to the left side of the plot, because this is the part of parameter space where the sbottom is also light.
As can be inferred from Fig.~\ref{fig:StopHiggsinoLimit}, changing the values of the higgsino mass in the $100-200\gev$ range do not significantly modify the structure of the bound. 

We do not consider here the case of a bino LSP for the reasons already explained above for $\tilde t_{R} \rightarrow \tilde B$ decays.

\subsection{Gluino Limits}
\label{subsec:gluino}

In this section, we add the gluino to the mix and consider the LHC limits, after $1\unit{fb}^{-1}$, on gluinos decaying through on or off-shell stops and sbottoms.  Recall from the discussion in Sect.~\ref{sec:NaturalPrimer} that the gluino mass is also important for naturalness because it corrects the Higgs potential at 2-loop order.  In this section, we will find that the gluino is constrained to be heavier than about $600-800\unit{GeV}$.  This means, from the point of view of naturalness, that the gluino mass limit is as important as the limits on stops discussed in Sect.~\ref{subsec:stop}.  

\begin{figure}[h!]
\begin{center} \includegraphics[width=1 \textwidth]{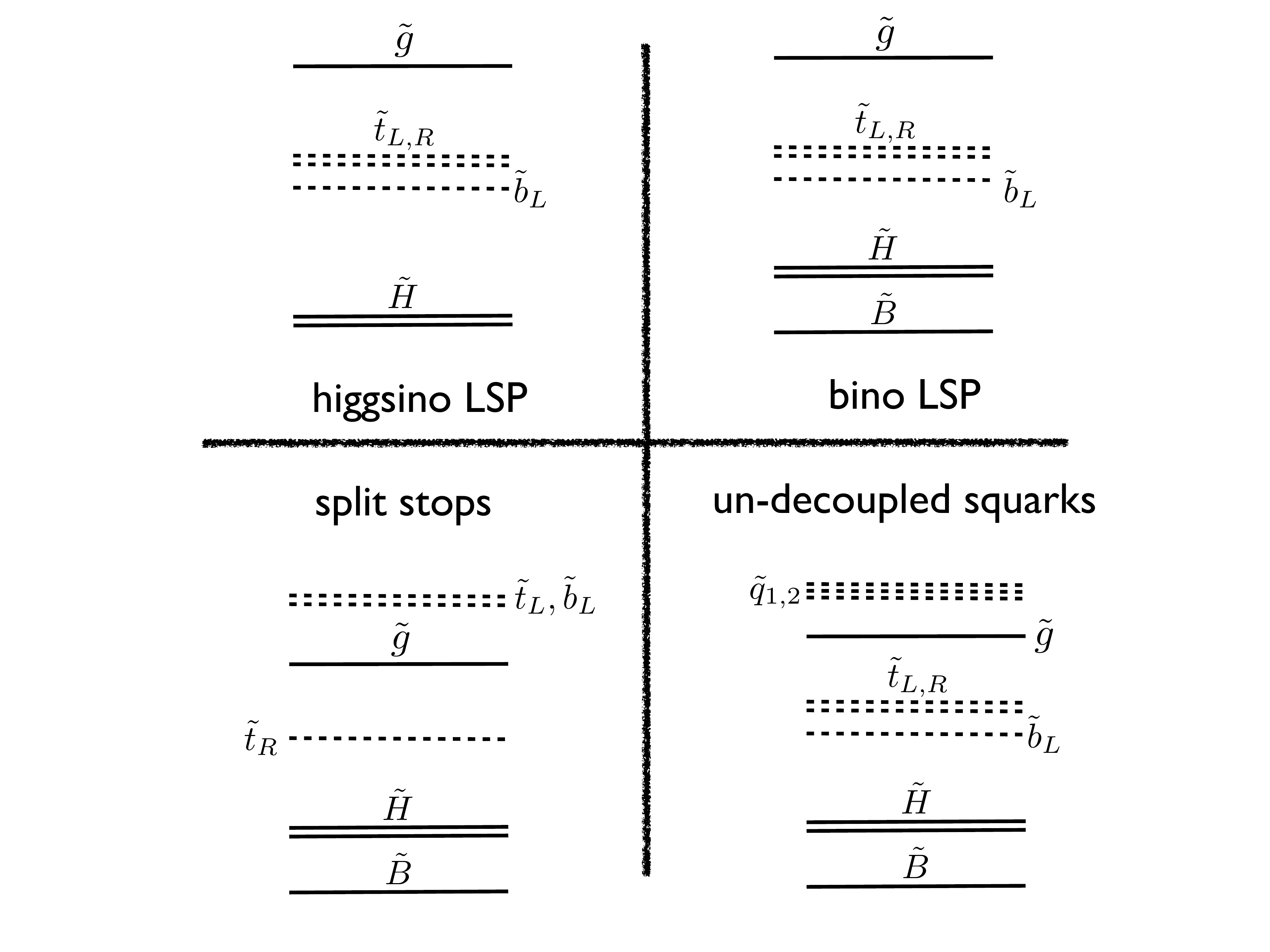} \end{center}
\caption{The four benchmark scenarios that we use to study limits on gluinos and stops.  In the {\bf higgsino LSP} scenario, we consider a gluino, degenerate stops and left-handed sbottom, and a higgsino LSP.  These are the minimal ingredients that need to be light for naturalness, and for simplicity we decouple the rest of the spectrum.  In the {\bf bino LSP} scenario, we add a bino with a soft mass of $M_1 = 100$~GeV.  In the {\bf split stops} scenario, we take the right-handed stop to be light and the left-handed stop/sbottom to be heavier than the gluino.  In the {\bf un-decoupled squarks} scenario, we test how the limit strengthens by lowering the mass of the first two generation squarks.
\label{fig:GluinoVarieties}
}
\end{figure}

We consider the limits on several different types of spectra, summarized in Fig.~\ref{fig:GluinoVarieties}, involving gluinos and light stops.  Throughout this section, for simplicity we neglect left-right stop mixing by taking $X_t = 0$. A non-zero $X_{t}$ will have minor effects on the region of parameter space where the bounds are dominated by gluino pair production, and will have the effect of weakening the bounds, to the levels already studied in the previous Section, when gluinos are too heavy to be relevant.

\begin{figure}[h!]
\begin{center} \includegraphics[width=1 \textwidth]{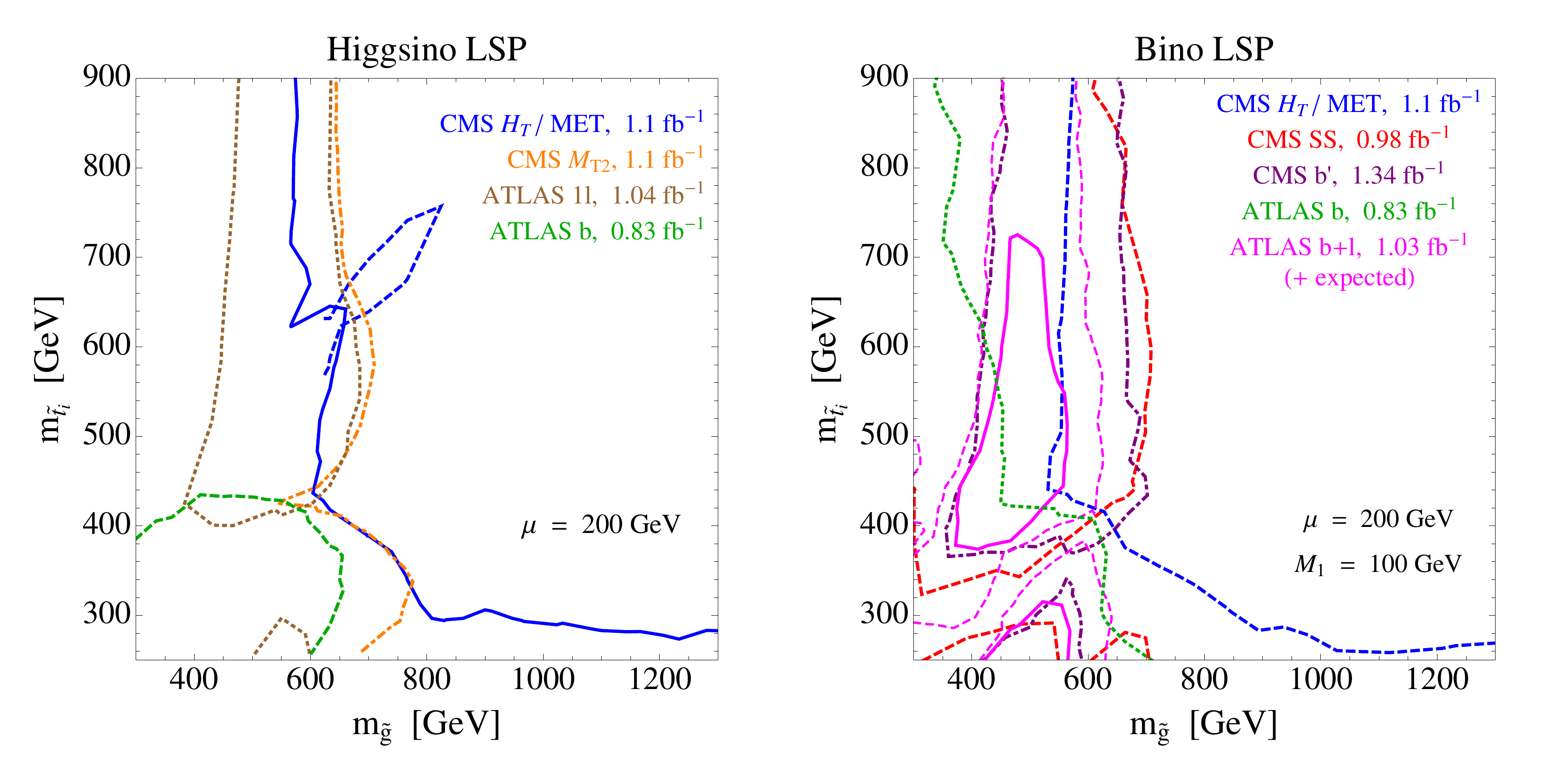} \end{center}
\caption{
\label{fig:StopGluinoLimit}
The limits on the {\bf Higgsino LSP} and {\bf bino LSP} scenarios, represented in terms of the gluino mass versus the degenerate stop pole masses.  In the limit of large gluino mass, we find that the strongest limit on direct stop/sbottom production, $m_{\tilde t} \gtrsim 300$~GeV, comes from searches for jets plus missing energy.   With only a higgsino LSP, the strongest limit on the gluino, $m_{\tilde g} \gtrsim 650$~GeV comes from searches for jets plus missing energy, and an ATLAS search for a single lepton plus jets and missing energy.  When both the bino and higgsino are light, we find that the strongest limit, $m_{\tilde g} \gtrsim 700$~GeV comes from the CMS search for same-sign dileptons plus missing energy.  To the left, the dashed blue line indicates a region of parameter space, $m_{\tilde t} \lesssim m_{\tilde g}$, that may also be excluded by the CMS search for jets plus missing energy.  However, the acceptance is highly sensitive to the precise value of the missing energy cut in this regime, signaling that the we cannot make a robust statement, given the precision of our simulation, in this part of parameter space. 
}
\end{figure}

{\bf Higgsino LSP.}  The first type of spectrum we consider, shown to the upper left of Fig.~\ref{fig:GluinoVarieties}, consists of a higgsino LSP, a light gluino, and light stops.  This spectrum constitutes the minimal ingredients that must be light for natural supersymmetry.  We choose to fix the higgsino mass to 200 GeV and vary separately the gluino mass and the mass of the stops, which we take to be degenerate, $m_{Q_3} = m_{u_3}$.  The limit that we find on this spectrum is shown to the left of Fig.~\ref{fig:StopGluinoLimit}.  For readability, we only show a selection of limit curves, including the searches that set the strongest limits.

In the high gluino mass region of the higgsino LSP parameter space, we find that the strongest limit comes from the CMS search for jets plus missing energy, $m_{\tilde t_i} \gtrsim 300$~GeV.  This is consistent with the limit we found on stops with a higgsino LSP in Fig.~\ref{fig:StopHiggsinoLimit} of Sect.~\ref{subsec:stop}, with the limit strengthened slightly because of the simultaneous presence of the left-handed stop/sbottom and the right-handed stop.  In the heavy stop part of the parameter space, we find that the strongest limit comes from the CMS $M_{T2}$ version of the jets plus missing energy search, and from the ATLAS search for 1 lepton plus jets and missing energy.  Here, the lepton comes from the decay of a top produced in the gluino decay, through an on or off-shell stop ($\tilde g \rightarrow t^+ t^- \tilde H^0$) or sbottom ($\tilde g \rightarrow t^\pm b^\pm \tilde H^\mp$).  We also find that the CMS search for jets plus missing energy may set the strongest limit along the line where the sbottom is slightly lighter than the gluino, $m_{\tilde b} \sim m_{\tilde t_i} \lesssim m_{\tilde g}$.  Here, the gluino decays to a soft b-jet plus a sbottom, which can decay to a very hard b-jet and a neutral higgsino, $\tilde b^\pm \rightarrow b^\pm \tilde H^0$.  The presence of two very hard jets in the final state leads to a high acceptance for the jets plus missing energy search.  However, we find that the acceptance in this regime is very sensitive to the precise value of the missing energy cut.  This prevents us from making a robust statement about the exclusion (hence the dashed line in the plot), after accounting for the uncertainties of our simulations.

{\bf Bino LSP.}  Second, we consider the limit on gluinos and stops with a bino LSP at 100 GeV and a higgsino at 200 GeV, as shown to the upper right of Fig.~\ref{fig:GluinoVarieties}.  One motivation for adding a bino is that it allows for mixed bino/higgsino dark matter. From the kinematics point of view, the interesting effect is that the bino lengthens the supersymmetric cascades.  Typically, the stops will decay first through the higgsinos (because the top Yukawa is stronger the hypercharge gauge coupling), which then decay to the bino, $\tilde H^\pm \rightarrow W^\pm \tilde B$, $\tilde H^0 \rightarrow Z \tilde B$, through the higgsino/bino mixing angle.  The limit on the bino LSP spectrum is shown to the right of Fig.~\ref{fig:StopHiggsinoLimit}.  Once again, searches for jets plus missing energy set the strongest limit on the stop mass, in the large gluino mass limit.  

The important difference between the bino and higgsino LSP scenarios is that the strongest limit on the gluino mass, $m_{\tilde g} \gtrsim 700\unit{GeV}$, comes from searches from same-sign dileptons plus missing energy.  There are two searches of this type conducted by CMS that set comparable limits, one supersymmetry search~\cite{CMS-PAS-SUS-11-010} and one search looking for pair production of $b'$ decaying to tops and $W$'s~\cite{CMS-PAS-EXO-11-036}.  The reason that same-sign dileptons become a powerful probe with the addition of the bino, is that leptons are produced both by the decays of tops and by the decays of leptonic $W$'s produced when the charged higgsino decays to the bino.  We find no limit from same-sign dileptons when the sbottom mass is lowered such that it can no longer decay to a top and a chargino,  $m_{\tilde b_L} \sim m_{\tilde t_i} < m_{\tilde H} + m_t$, reducing the number of leptons in the final state.  As the stop/sbottom mass is further lowered, the limit is recovered because $\tilde g \rightarrow \tilde t_i^\pm t^\mp$ opens up.  The result, in our parameter space, is a gap in same-sign coverage from $m_{\tilde t_i} \sim m_{\tilde b_l} \approx 300-400\unit{GeV}$.  Our choice of $\mu$ changes the position of this gap, but does not affect the overall limit since the search for jets plus missing energy covers this gap and sets the strongest limit in this regime.

{\bf A somewhat squashed spectrum.}  Next, we deform the bino LSP spectrum by squashing the mass splitting between the gluinos and the higgsino/bino.  Compressing the spectrum has the impact of reducing the amount of visible and missing energy, typically resulting in weaker limits on superpartner masses~\cite{SquashRefs}.  However, it should be kept in mind that the compression itself may be a new form of tuning (in the form of a relation between the colored superpartner and LSP mass) depending on the UV completion, therefore it is not totally clear whether or not compressed MSSM spectra are really more natural (extending the field content beyond the MSSM, small mass splittings can occur naturally, see for example~\cite{stealth}).  

\begin{figure}[h!]
\begin{center} \includegraphics[width=1 \textwidth]{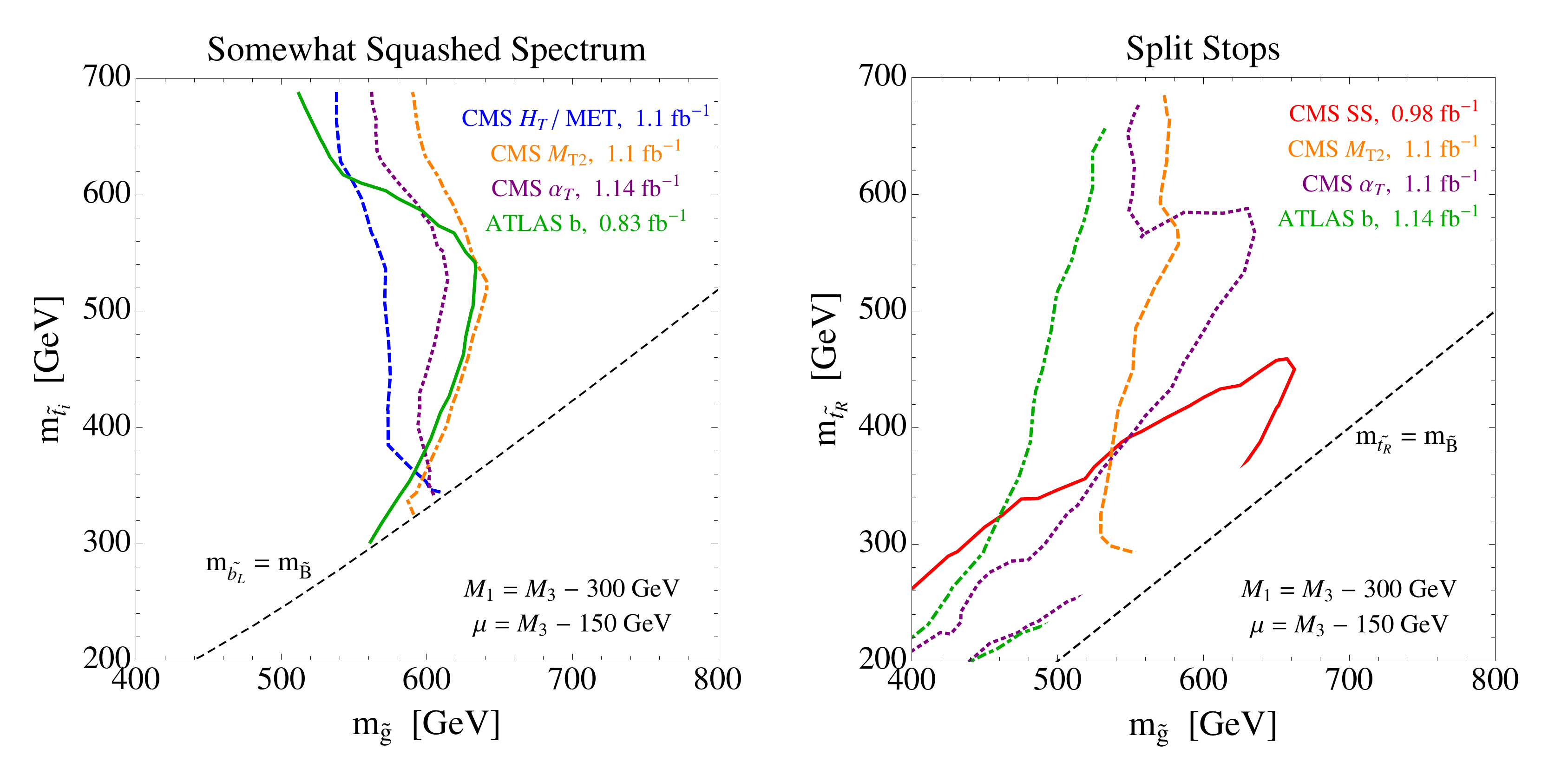} \end{center}
\caption{
\label{fig:OtherStopGluinoLimit}
Here we show how the gluino versus stop mass limit changes when the spectrum is compressed ({\it left}), or when the stop masses are split ({\it right}).  For the compressed case, we modify the {\bf bino LSP} benchmark by fixing the mass splitting between the gluino and the LSP to be moderately compressed, $M_3 - M_1=300\unit{GeV}$, and the limit on the gluino weakens to $m_{\tilde g} \gtrsim 550-600$~GeV.  For the {\bf split stops} scenario, the left handed stop/sbottom are taken heavier than the gluino.  The mass of the right-handed stop determines which search dominates the gluino mass limit.  Same-sign dileptons set the strongest limit when $\tilde g \rightarrow t \, \tilde N_i$ is kinematically allowed.  For heavier stops, the dominant gluino decay is the one-loop decay $\tilde g \rightarrow g \, \tilde N_i$, and the strongest limit comes from jets plus missing energy.
}
\end{figure}

In the previous case, we fixed the bino and higgsino masses to 100 and 200 GeV, respectively, while varying the gluino and stop masses.  Now, we hold constant the splitting between the gluino mass and the bino/higgsino, choosing $M_1 = M_3 -300\unit{GeV}$ and $\mu = M_3 - 150\unit{GeV}$.  The resulting limits are shown to the left of Fig.~\ref{fig:OtherStopGluinoLimit}.  The compression has the effect of squeezing out many of the decay modes involving tops, for example the three-body decays $\tilde g \rightarrow t^- t^+ \tilde H^0 (\tilde B)$ are now kinematically disallowed.  This reduces the number of leptons in the final state, and the strongest limits on the gluino mass, $m_{\tilde g} \gtrsim 600$~GeV, now come from searches for jets (with and without b-jets) and missing energy.

\begin{figure}[h!]
\begin{tabular}{S S}
\includegraphics[width=0.5 \textwidth]{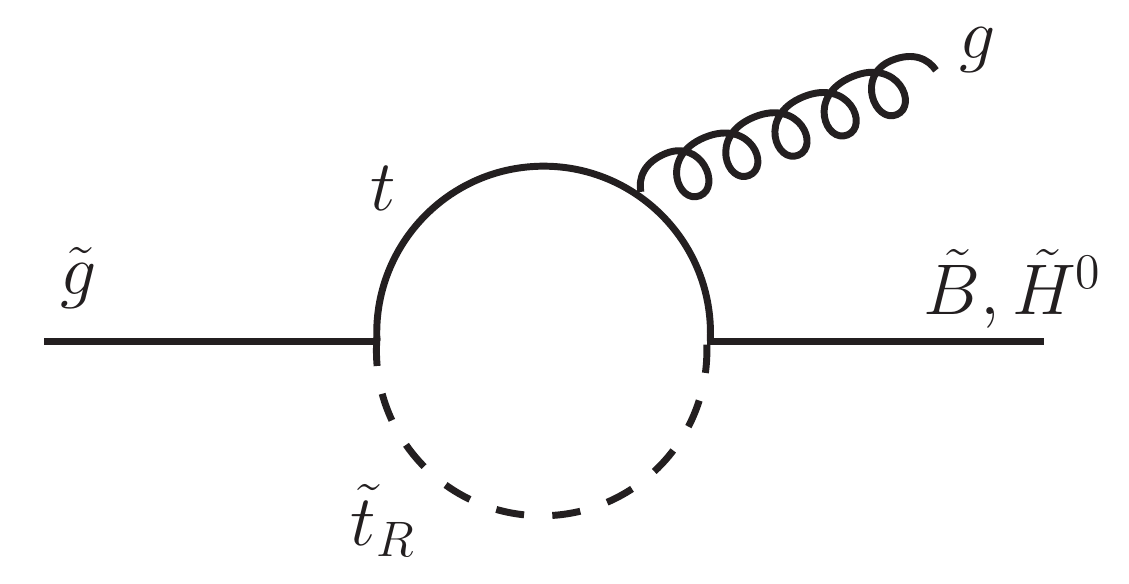} &  \includegraphics[width=0.5 \textwidth]{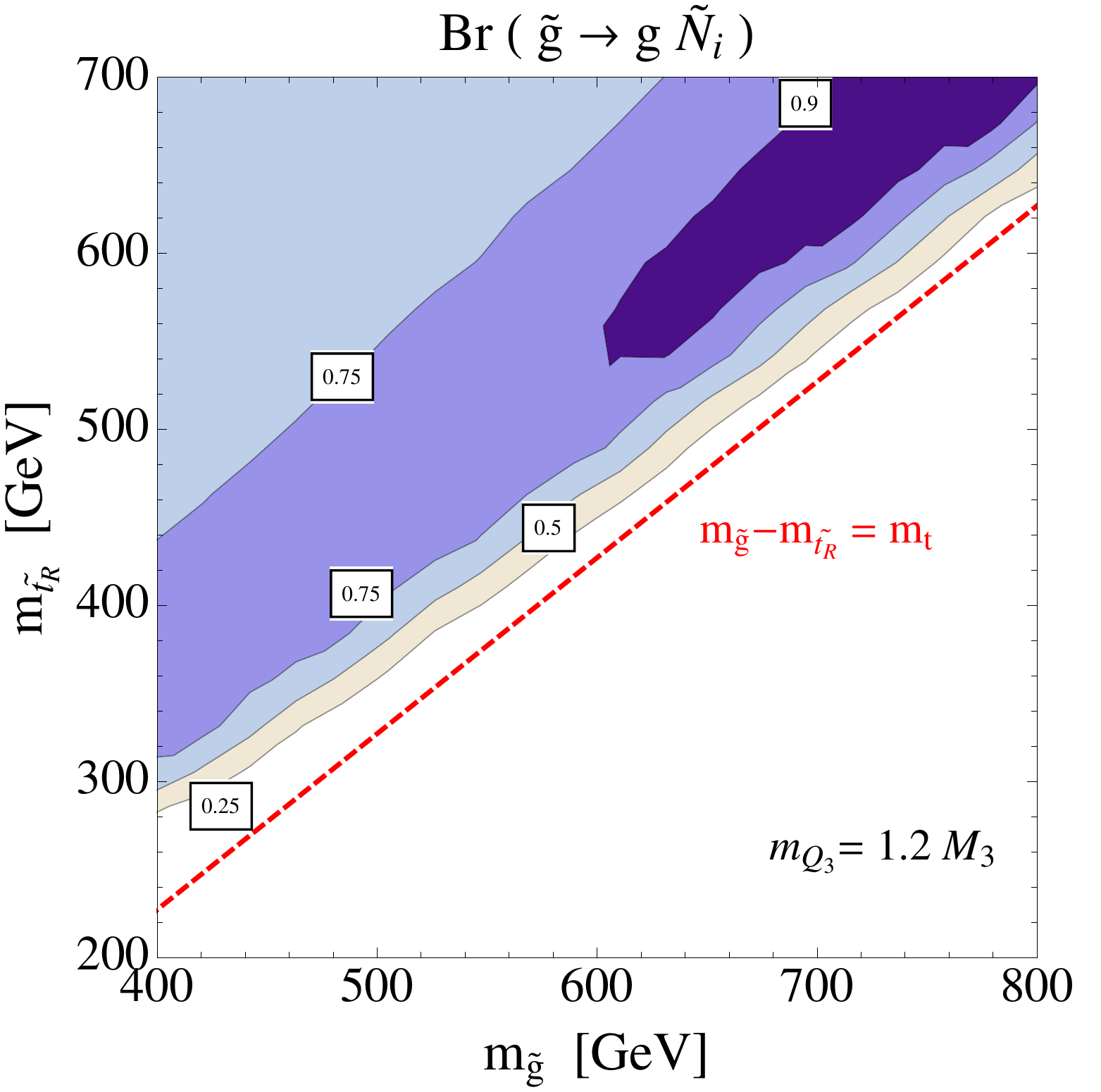} \\
\end{tabular}
\caption{\label{fig:OneLoop}
The dominant decay of the gluino can be the one-loop diagram shown to the {\it left}, $\tilde g \rightarrow g \, \tilde N_1$, with the stop running in the loop.  The branching ratio of this decay path is shown to the {\it right} within the parameter space of our {\bf split stops} benchmark scenario.  This decay dominates when the right-handed stop is heavy enough to close the two body decay to a top, $\tilde g \rightarrow t \tilde N_i$, as long as the other squarks are sufficiently heavy to suppress competing three-body decays.  For this example, we have taken $m_{Q_3} = 1.2 \, M_3$ which is sufficient to suppress the three-body decay mediated by the sbottom, $\tilde g \rightarrow b b \tilde N_1$, relative to the one-loop decay.
}
\end{figure}

{\bf Split Stops.} We now consider the effect on the gluino mass limit when the stop masses are no longer degenerate, as shown in the lower left of Fig.~\ref{fig:GluinoVarieties}.  We vary the gluino mass and the right-handed stop mass, keeping the left-handed stop/sbottom heavier than the gluino, $m_{Q_3} = 1.2 \,M_3$.  While this choice is less justified by naturalness arguments,  it is an interesting case to consider because it highlights different final states with different kinematics.  The bino and higgsino masses are chosen, as in the squashed spectrum considered above, to track the gluino mass,  $M_1 = M_3 -300\unit{GeV}$ and $\mu = M_3 - 150\unit{GeV}$.  The most interesting feature of the split stop case is that, when the two-body decay of the gluino to the stop and a top is kinematically forbidden, $m_{\tilde t_R} > m_{\tilde g} - m_t$, the gluino dominantly decays through a top/stop loop to a gluon and a neutral higgsino or bino, $\tilde g \rightarrow g \, \tilde H^0 (\tilde B)$~\cite{Barbieri:1987ed}, as shown in the Feynman diagram to the left of Fig.~\ref{fig:OneLoop}.  We have used the program SDECAY~\cite{Djouadi:2006bz} to compute the branching ratio of this decay in the parameter space we consider, and we find that it typically dominates over the three-body decay through the off-shell sbottom, $\tilde g \rightarrow b^+ b^- \tilde H^0 (\tilde B)$, as shown to the right of Fig.~\ref{fig:OneLoop}.

The limit on the gluino mass, with split stops, is shown to the right of Fig.~\ref{fig:OtherStopGluinoLimit}.  There are two important regimes, depending on whether or not the two body decay of the gluino to a top and the stop is open.  When $m_{\tilde t_R} < m_{\tilde g} - m_t$, every event contains four tops, and we find that same-sign dileptons set the strongest limit, with the leptons coming from top decays.  When $m_{\tilde t_R} > m_{\tilde g} - m_t$, the one-loop gluino decay dominates, as discussed above, and the strongest limit comes from the CMS $\alpha_T$ version of the search for jets and missing energy.  Further raising the stop mass, the three body decay to bottoms becomes competitive with the one-loop decay, $\tilde g \rightarrow b^+ b^- \tilde H^0 (\tilde B)$, and the strongest limit comes from a channel of the CMS $M_{T2}$ search that demands 1 b-jet.

\begin{figure}[h!]
\begin{center} \includegraphics[width=0.5 \textwidth]{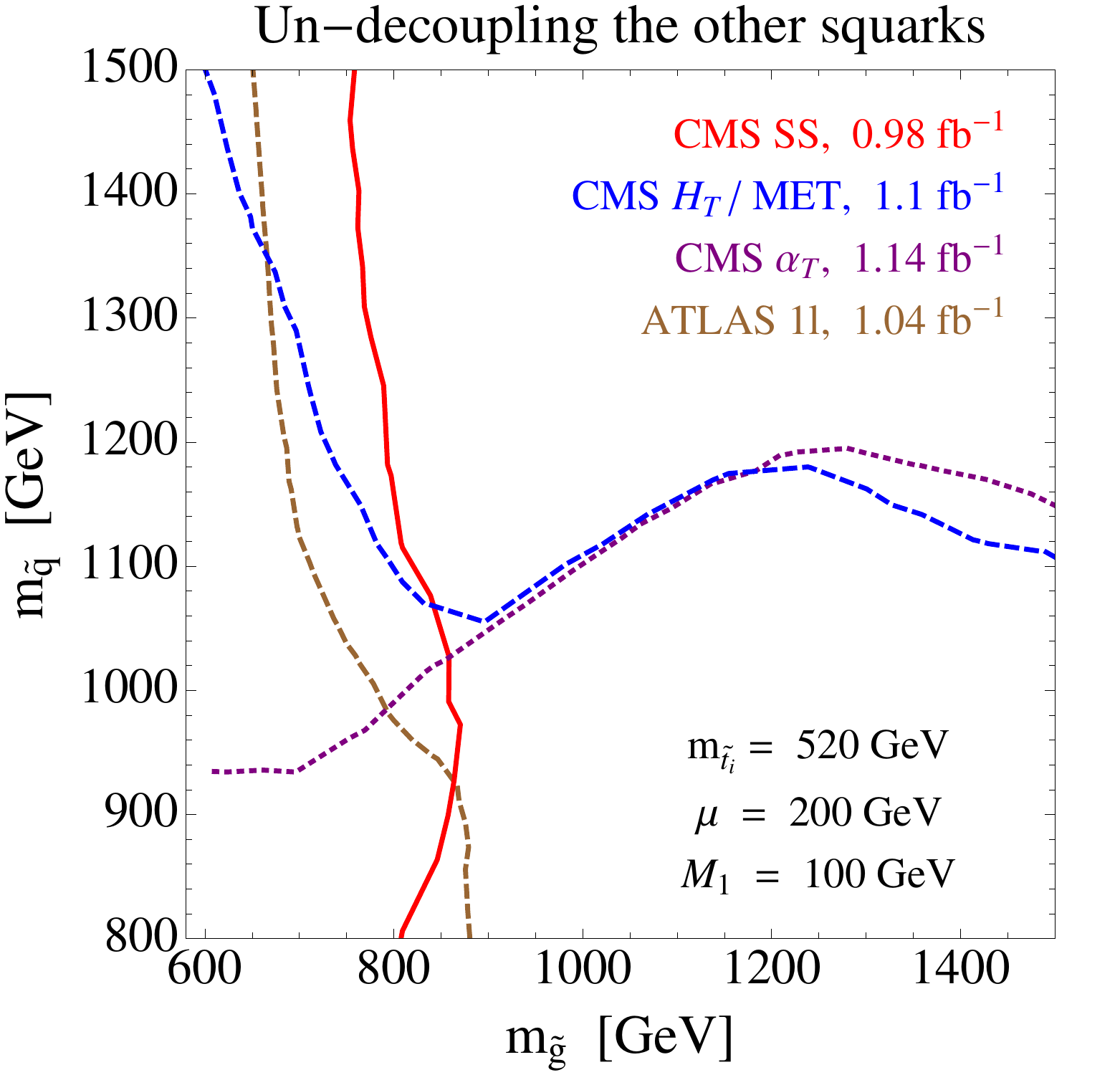} \end{center}
\caption{
\label{fig:Undecouple}
The limit on the gluino mass versus a common mass for the squarks of the first two generations in the {\bf un-decoupled squark} benchmark.  We find that searches for jets plus missing energy demand that $m_{\tilde q} \gtrsim 1.1\unit{TeV}$, and above this mass the effect of the extra squarks on the gluino limit quickly decouples.
}
\end{figure}

{\bf Un-decoupled Squarks.}  So far, in all of the above benchmarks, we have decoupled the squarks of the first two generations.  This choice was motivated by naturalness, since the limits on the gluino and stops are weaker when the squarks of the first two generations are heavy.  We conclude our discussion of the limits on gluinos by testing exactly how heavy the squarks need to be.  We answer this question by deforming the bino LSP benchmark, as shown to the lower right of Fig.~\ref{fig:GluinoVarieties}.  We vary the gluino mass against a common mass chosen for all of the squarks of the first two generations, $m_{\tilde q} = m_{Q_{1,2}} = m_{u_{1,2}}  =m_{d_{1,2}}$.  We fix both stop soft masses to 520 GeV and, as above, we choose $M_1=100\unit{GeV}$ and $\mu = 200\unit{GeV}$.  The limit on this scenario is shown in Fig.~\ref{fig:Undecouple}.  In the limit of heavy gluino mass, the strongest constraint comes from searches for jets and missing energy, and the common squark mass must be heavier than about 1 TeV.  The strongest limit on the gluino mass comes from same-sign dileptons, as in Fig.~\ref{fig:StopGluinoLimit}.  As the squark masses are raised, they very quickly decouple, and have little effect on the gluino mass once $m_{\tilde q} \gtrsim 1.2\unit{TeV}$.

\section{Implications for SUSY Models}
\label{sec:interpretation}

In this section we briefly consider some implications of our results for various SUSY models.  We discuss the interplay of LHC results with the LEP-2 bound on the Higgs mass in the MSSM, the consequences of the LHC limits for the flavor structure of the squark soft masses, and finally we will also consider the limit on natural spectra with gaugino unification.

\begin{figure}[h!]
\begin{center} \includegraphics[width=1 \textwidth]{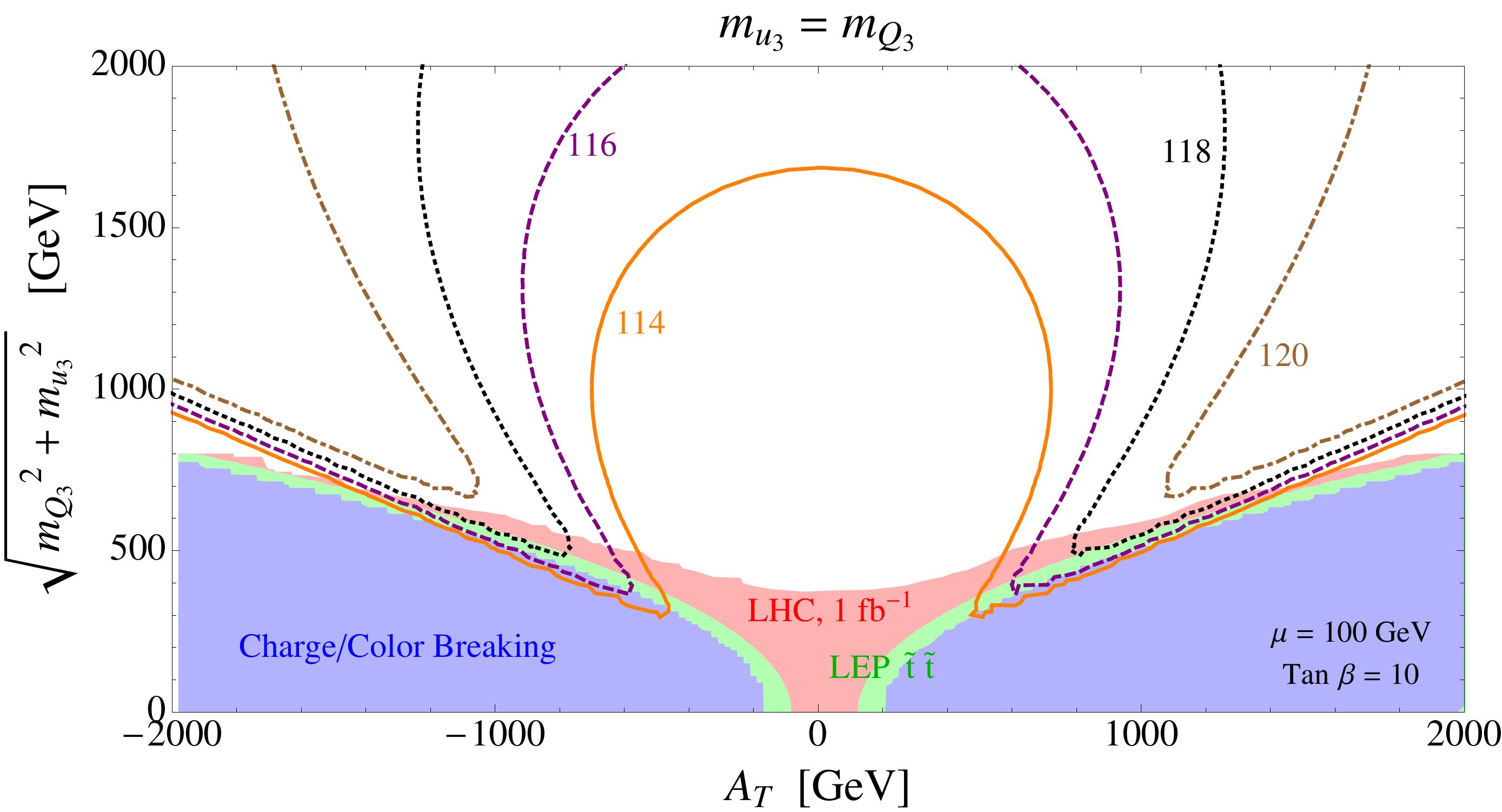} \end{center}
\caption{\label{fig:HiggsMass}
Here we show the interplay of the LHC limits that we have found on the stops and left-handed sbottom with the LEP-2 limit on the Higgs mass.  We specialize to higgsino LSP, with $\mu = 100$~GeV.  We vary the stop $A$-term and the square root of the average stop soft mass squared.  This unconventional parameterization emphasizes the fine-tuning of the electroweak sector, which, as we discuss in the text, corresponds to the \emph{squared} distance from the origin of the plot.  The red shaded region is the exclusion we find from LHC searches for jets plus missing energy.  The green region corresponds to a stop lighter than 100 GeV and is excluded by LEP-2.  In the blue region, large left-right stop mixing leads to a tachyonic stop and charge/color breaking.  The Higgs mass contours emphasize that the LHC is now beginning to probe the region allowed by the LEP-2 Higgs mass exclusion, increasing the fine-tuning in the MSSM.}
\end{figure}

We begin this section by discussing how the LHC limits relate to the LEP-2 bound on the Higgs mass.  As we stressed in the introduction and in Sect.~\ref{sec:NaturalPrimer}, there are two logically different reasons why the MSSM may need to be finely-tuned.  The first is the little hierarchy problem which results from the LEP-2 limit on the Higgs mass, and the second is the new set of LHC limits on those superpartners that are relevant for naturalness, like the stops.  So far in this paper, we have focused only on the direct limits without any concern for the Higgs mass, because the little hierarchy problem is model dependent and can be alleviated by modifications to the Higgs sector of the MSSM, which may or may not substantially affect the stops and gluino phenomenology.  However, it is interesting to ask how these two sources of fine-tuning are related without extending the MSSM. The answer to this question is shown in Fig.~\ref{fig:HiggsMass}, where we present both the LHC stop limit, derived from our simulations, and the contours of constant Higgs mass, using the one-loop renormalization improved result of~\cite{Haber:1996fp}.  This plot corresponds to higgsino LSP with $\mu = 100\unit{GeV}$, $\tan \beta = 10$, and degenerate stop soft masses, $m_{u_3} = m_{Q_3}$.  We also show the region that is excluded by LEP-2 because one of the stops is lighter than about 100 GeV, and the region where one of the stops becomes tachyonic, due to large left-right stop mixing, leading to charge and color breaking.

We have chosen, in Fig.~\ref{fig:HiggsMass}, to represent the LHC stop limit, and the Higgs mass contour, in a plane parameterized by the stop $A$-term and by the square root of the average of the left/right stop soft masses squared, $\sqrt{m_{Q_3}^2 + m_{u_3}^2}$.  In this parameterization (thanks to Pythagoras) the fine-tuning of the electroweak sector is simply the \emph{square} of the linear distance from the origin, as can be easily seen by examining equation~\ref{eq:der1}. We note immediately, by inspecting Fig.~\ref{fig:HiggsMass}, that, prior to the LHC, the region of the MSSM with the least fine-tuning was the so-called ``maximal mixing" scenario, where $X_t \sim A_t = \sqrt{6} m_{\tilde t}$, because this is where the $m_h = 114\unit{GeV}$ contour passes closest to the origin.  We find that this region of the plot is now becoming excluded by LHC searches, showing that there is a complementarity between the LHC limits and the LEP-2 limit on the Higgs mass.  In other words, the LHC is now beginning to make the fine-tuning worse in the MSSM.  Or more positively stated, the LHC is starting to probe the most interesting part of parameter space that remains in the MSSM.
While this statement at the moment strongly depends on having higgsinos lighter than stops (which is still not absolutely required by naturalness arguments), these results are likely to become more robust in the next months.

\begin{figure}[h!]
\begin{center} \includegraphics[width=1 \textwidth]{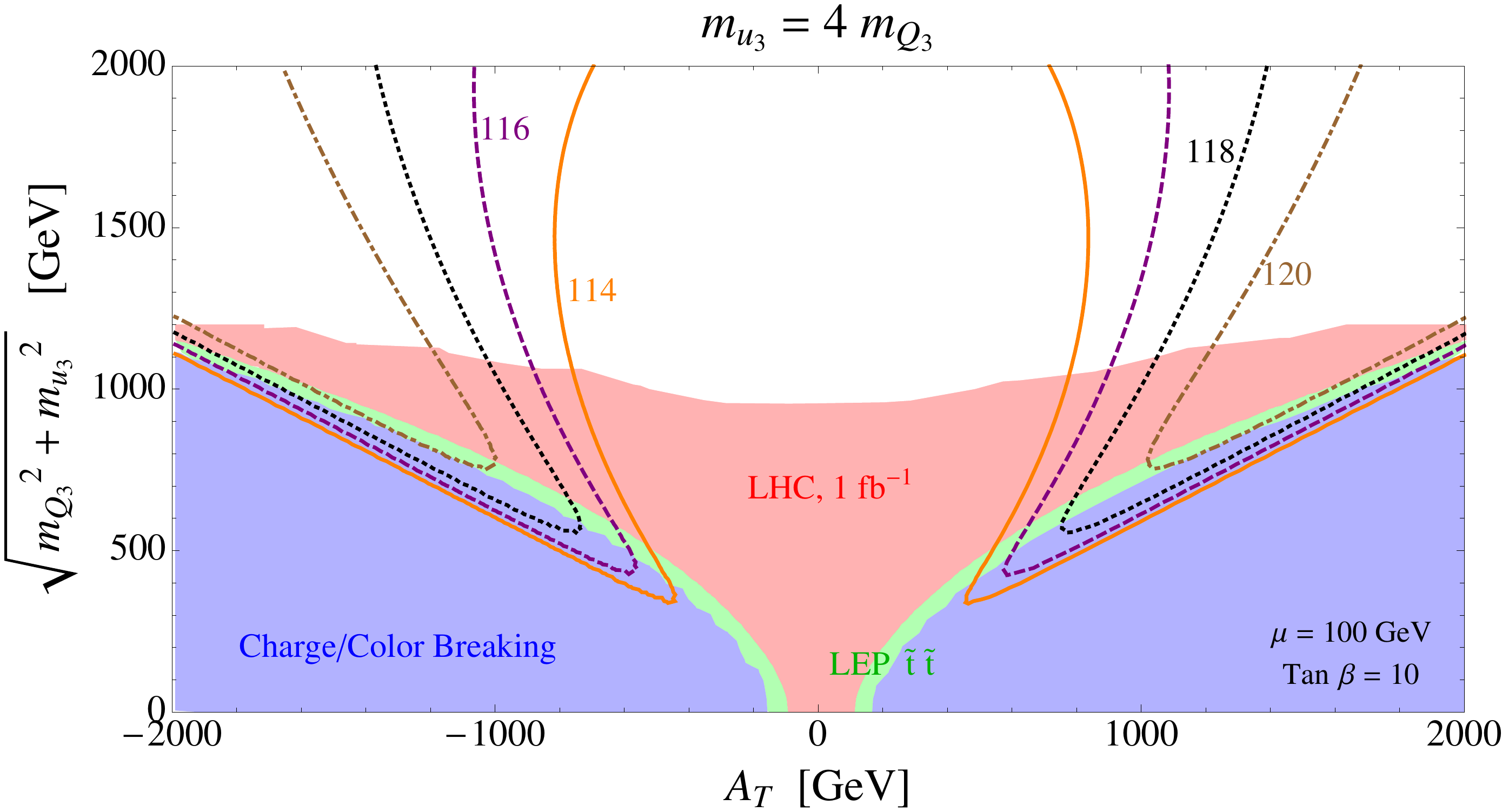} \end{center}
\caption{\label{fig:HiggsMassSplit}
The same as Fig.~\ref{fig:HiggsMass}, except instead of taking the left/right stop soft masses degenerate, as above, we fix $m_{u_3} = 4 \, m_{Q_3}$.  This has the impact of increasing the region of the plot that is excluded by the LHC,  which sets a limit on the lighter stop and sbottom, whose masses are determined here by $m_{Q_3}$.  Meanwhile, the fine-tuning (the y-axis scale) and the radiative contribution to the Higgs mass are driven by the heavier stop, determined here by $m_{u_3}$.  The difference between this figure and Fig.~\ref{fig:HiggsMass} highlights why naturalness prefers the situation where both stops are roughly degenerate.
}
\end{figure}

We also show, in Fig.~\ref{fig:HiggsMassSplit}, what happens when the stop soft masses are non-degenerate, by fixing $m_{u_3} = 4 \, m_{Q_3}$.  In this case, the LHC carves out a larger region of the parameter space where the Higgs mass satisfies the LEP-2 limit.  This behavior can be understood simply.  The LHC primarily limits the lightest stop (and the sbottom), whose masses in this case are determined by $m_{Q_3}$.  On the other hand, the radiative contribution to the Higgs mass, and the fine-tuning which determines the position on the y-axis, is primarily driven by the largest stop soft mass, here $m_{u_3}$.  The result is that the LHC limit is stronger in the interesting part of parameter space.  By comparing figures~\ref{fig:HiggsMass} and~\ref{fig:HiggsMassSplit}, we see that naturalness prefers spectra where the two stop soft masses are comparable, $m_{u_3} \sim m_{Q_3}$.

\begin{figure}[h!]
\begin{center} \includegraphics[width=1 \textwidth]{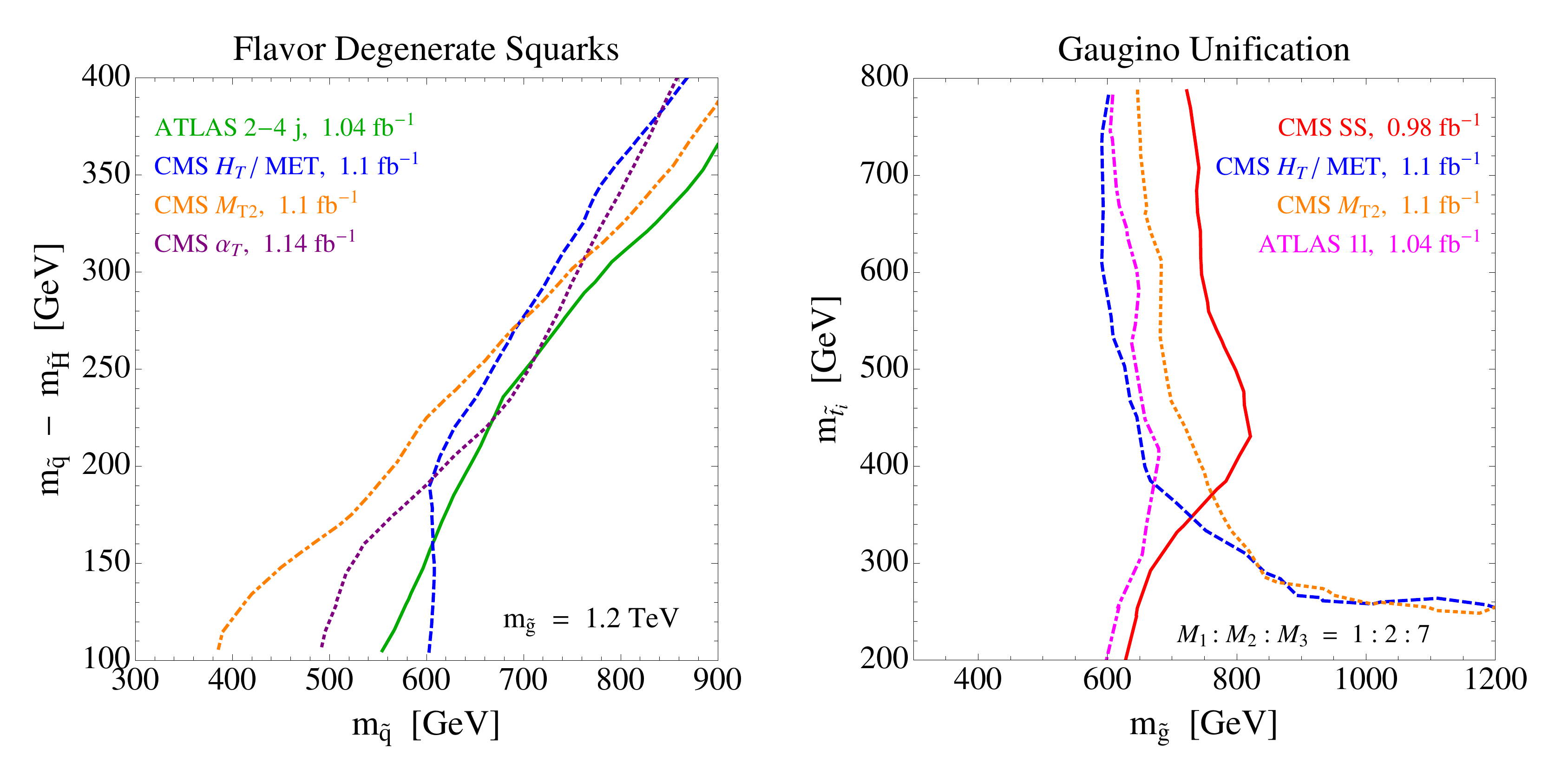} \end{center}
\caption{\label{fig:Unify}
On the {\it left} we show the limit when the squarks have flavor universal masses and the higgsino is the LSP\@.  We have fixed the gluino mass to 1.2 TeV and we vary the common squark mass and the mass splitting between the squarks and the higgsino.  We see that if the spectrum is compressed, the squarks can be as light as 600 GeV, with the strongest limit coming from searches for jets and missing energy.  This represents a sort of ``best case scenario" for flavor degeneracy because the fine-tuning (both in the compression and the electroweak symmetry breaking) is only moderate.  On the {\it right} we show the limit on gluino versus stop mass, imposing gaugino unification, $M_1 :  M_2 : M_3 \approx 1:2:7$.  We consider degenerate stops, with the first two generation squarks decoupled.  We find that the gluino is constrained to be heavier than about 750-800 GeV, with the strongest limits coming from same-sign dileptons plus missing energy and jets plus missing energy.}
\end{figure}

Next we consider the implication of the LHC limits for the flavor structure of the squark soft masses.  Since fine-tuning is determined by the stop soft masses, while the strongest limits are on the light squarks, the obvious way to reduce fine-tuning is to consider spectra with a flavor non-degenerate squark soft mass, so that the stops are lighter than the squarks of the first two generations.  This scenario has been the focus of our limit study in Sect.~\ref{sec:limits}.  However, as pointed out in Sect.~\ref{sec:status}, the flavor degenerate case for the squarks may not be strongly disfavored yet, due to the dependence of the LHC constraints on the LSP mass. Therefore, it is also interesting to consider flavor degenerate squarks (which are predicted by many of the simplest scenarios of SUSY breaking, such as gauge mediation), and to check how strong the limits really are. This is the subject of the left side of Fig.~\ref{fig:Unify}, where we show the LHC limit coming from the scenario where all squarks are flavor degenerate at the electroweak scale (including stops and sbottoms), and the gluino mass is fixed to 1.2 TeV, which is heavy enough to deplete the rate of associated gluino-squark production.  
Here we also made the simplifying choice of taking the $Q,U,D$ soft masses to be equal, although moderate splittings do not drastically change our conclusions.

We consider a higgsino LSP and separately scan the common squark mass and the squark-higgsino mass splitting.  We see that if the spectrum is mildly compressed, with a squark-higgsino splitting varying from 100-250 GeV, then the limit on the squark masses is in the 600 to 700 GeV ballpark range.  This limit (and also the 1.2 TeV gluino) corresponds to about 10\% fine-tuning in the Higgs potential, which represents a ``best case" scenario for a flavor degenerate boundary condition. 

It is also likely that the flavor-degenerate option will be more easily constrained by the future releases of the LHC data (unless, of course, a signal is found) and may be disfavored in the next months. If this will be indeed the case, in the context of R-parity conserving natural SUSY models with MSSM-like signatures, one is naturally led to consider ``flavorful'' SUSY breaking scenarios where the third generation squarks is split from the first two generation already at the SUS mediation scale. The investigation of such models is not new~\cite{decouple1st2nd} and was initially motivated by flavor considerations. 

Not that the flavor non-universal contribution to the squark mass matrices should be at least of the same order or larger that the flavor-blind one. Generically, if the SUSY mediation mediation mechanism does not commute with flavor, it is likely that additional sources of flavor violation beyond the Minimal Flavor Violation~\cite{mfv} are introduced, as confirmed in explicit model constructions~\cite{u2studies}. These new sources of flavor violation may be detectable in experiments, such as LHCb or a future SuperB factory~\cite{bphysreach}, providing an interesting complementarity between direct and indirect searches. 

However this is not necessarily the case if one can ensure that, even after including the SUSY breaking and mediation sectors, the SM Yukawa couplings are the only sources of flavor breaking. One possible way to achieve this result could be to couple the SUSY breaking sector directly to the SSM Higgs sector and hence use the Yukawa couplings to transmit to the squark soft mass matrices a flavorful SUSY breaking contribution, from an initially flavor-blind SUSY breaking sector~\cite{higgsmodels}.

We conclude this section with a brief discussion of the limit on gaugino unification.  Recall that throughout this paper, we have decoupled the superpartners whose masses are inconsequential for naturalness, including the bino and the wino.  But it is also interesting to relax this assumption and consider spectra where both the bino and wino are light, because many models of supersymmetry breaking, with gauge coupling unification, predict that the gaugino masses appear in the ratio\footnote{For brevity we do not explicitly consider other gaugino mass relations, such as the anomaly-mediated one, $M_{1} : M_{2} : M_{3} \approx 3.3 : 1 : 9$, since from kinematical considerations the limits should not be very different that those presented here.} $M_1 : M_2 : M_3 \approx 1 : 2 :7$.  Naturalness constrains the gluino to be light, and then, if gaugino unification holds, the wino and the bino should also be light.  We show the limit on natural supersymmetry with gaugino unification in Fig.~\ref{fig:Unify}, where we separately vary the stop masses  and the gluino mass, while  fixing the bino and wino masses to satisfy the gaugino unification relation.  The stops are taken to be degenerate, with no left-right mixing, and the squarks of the first two generations are decoupled to 3 TeV.  The presence of both the bino and wino has the effect of lengthening the supersymmetry cascades, similarly to the bino LSP scenario that we considered in Sect.~\ref{subsec:gluino}.  The limit is similar to that case, with the gluino constrained to be heavier than about 800 GeV by the search for same-sign dileptons plus missing energy.  As in the other cases, the stops and the left handed sbottom are constrained, when degenerate, to be heavier than about 250 GeV by searches for jets plus missing energy.

\begin{figure}[h!]
\label{fig:StopHiggsinoReach}
\begin{center} \includegraphics[width=1 \textwidth]{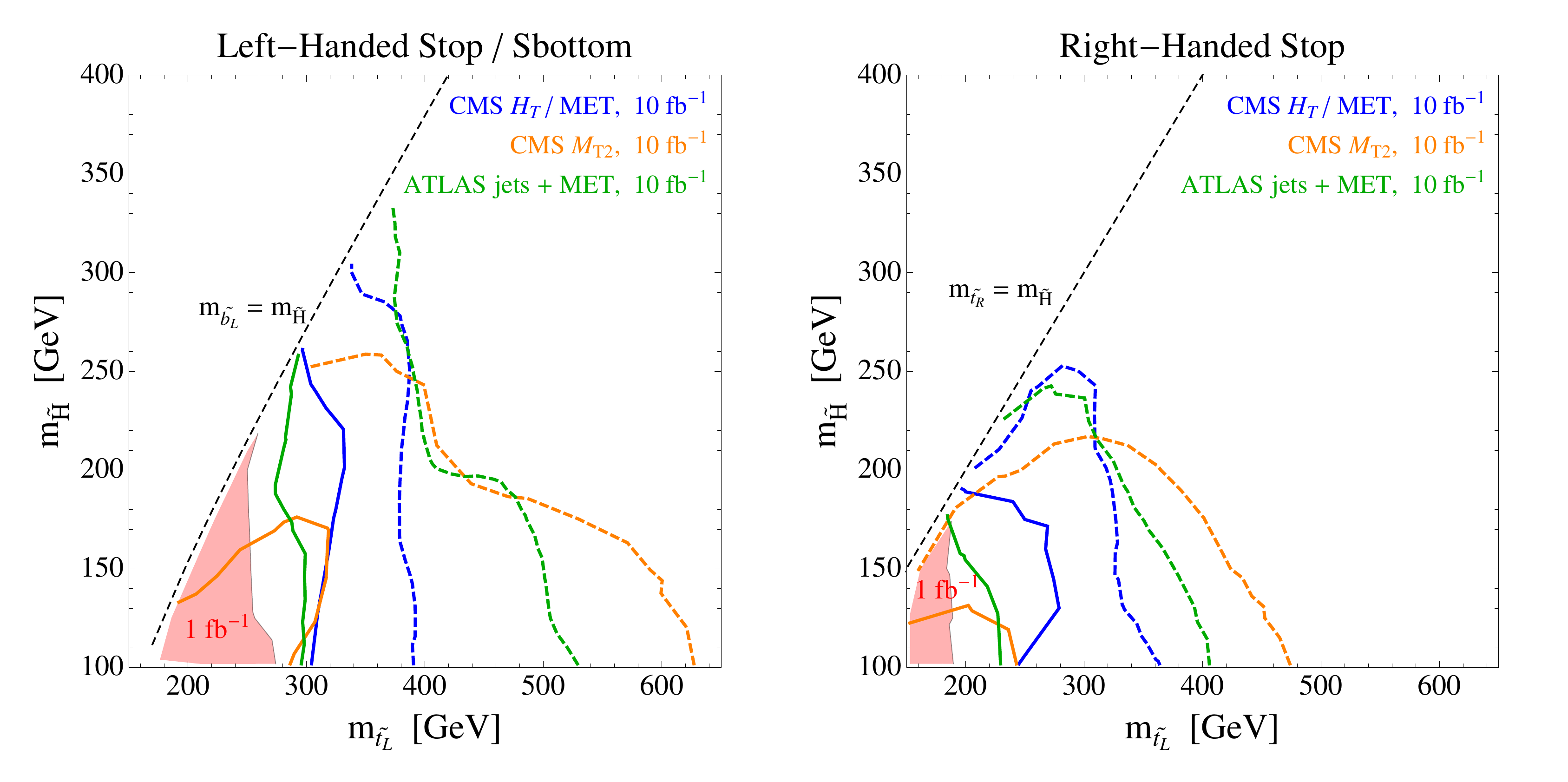} \end{center}
\caption{The estimated 95\% exclusion reach, with $10\unit{fb}^{-1}$, for left-handed stop/sbottom ({\it left}) and right-handed stop ({\it right}), with higgsino LSP.  We show the reach by extrapolating the cuts of the existing searches for jets and missing energy.  We find that the reach is highly sensitive to the treatment of systematic errors.  For the solid curves, we assume that statistical errors will reduce with luminosity but that systematic errors will remain a constant fraction of the background estimate.  For the dashed curves, we take the idealized limit of zero statistical or systematic uncertainties on the background estimate, taking the central value of the backgrounds reported in the current experimental searches. }
\end{figure}

\begin{figure}[h!]
\label{fig:GluinoStopReach}
\begin{center} \includegraphics[width=0.5 \textwidth]{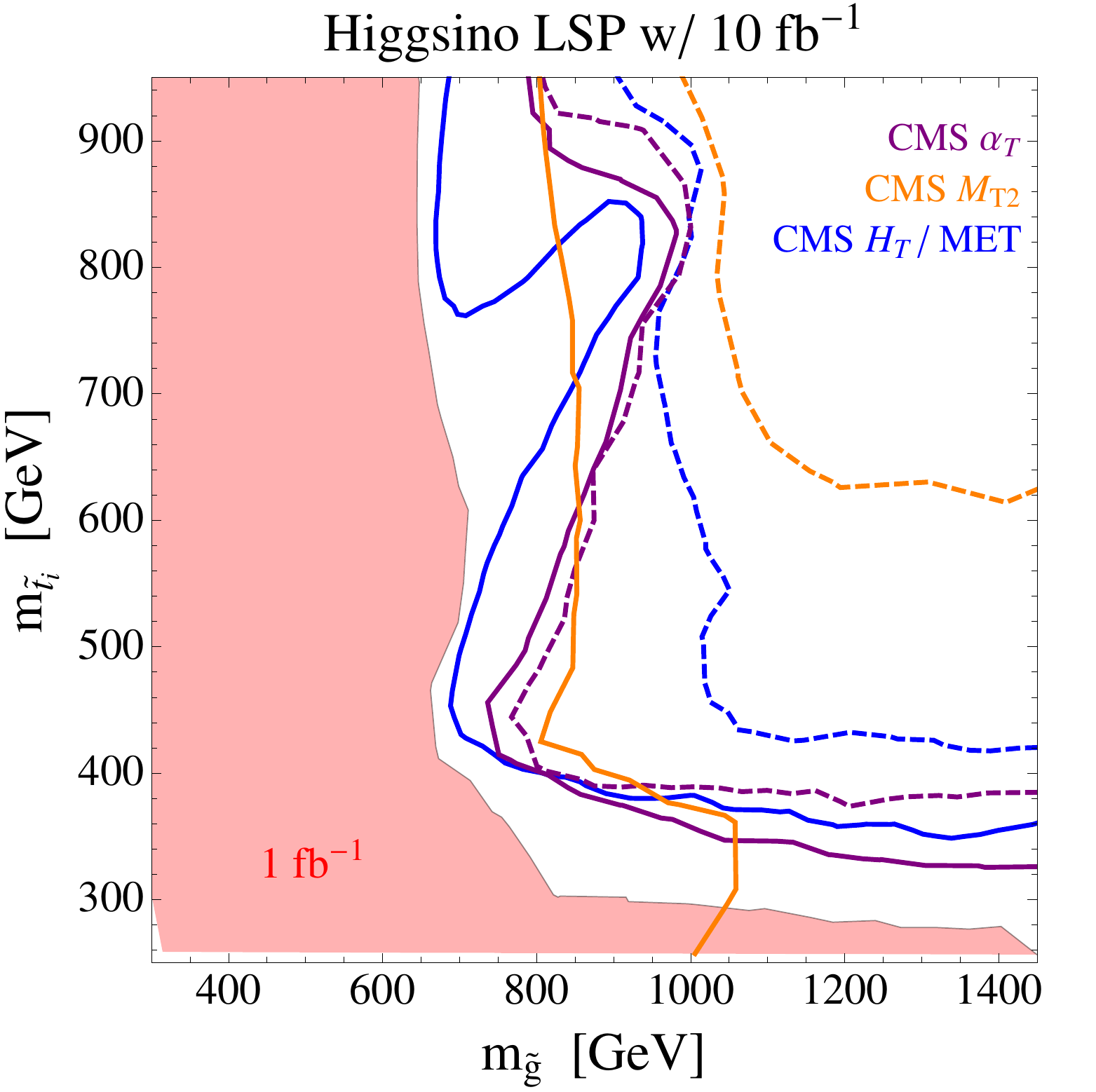} \end{center}
\caption{The estimated 95\% exclusion reach, with $10\unit{fb}^{-1}$, for the {\bf higgsino LSP} benchmark.  As in Fig.~\ref{fig:StopHiggsinoReach}, the solid lines extrapolate the current systematic and statistical errors on the background, while the dashed lines assume perfect knowledge of the background.  The large spread between these estimates emphasizes the importance of the eventual systematic errors for the reach.}
\end{figure}

\section{Conclusions}

We have investigated the current LHC limits on Natural SUSY, \ie~on supersymmetric scenarios where the higgsinos, the top squarks, the left-handed bottom squark, and the gluino are bound to be light from the requirement of natural electroweak symmetry breaking. We found that the most constraining searches are those looking for the jets+$\met$ signatures in the case of stops and sbottom decaying to neutralinos and charginos, while a combination of jets+$\met$ and same-sign (SS) dilepton searches for the cascades initiated by the gluino. Our main results are summarized in in Tables~\ref{tab:StopResults} and~\ref{tab:GluinoResults}, where we show the mass limits found for the various simplified models studied, together with a reference to the relevant plot in this paper.
The luminosity of $1fb^{-1}$ marks a divide in the LHC SUSY searches, after which it is possible to start looking in detail for direct production of third generation squarks, complementing the searches already looking for them in processes initiated by gluino pair production. With higher luminosities it will also be possible to probe direct higgsino production, which will be another necessary step towards probing the natural region of SUSY.

 \begin{table}[h!]
\begin{center}
\begin{tabular}{|c|c||c|c|}
\hline
production & LSP & $\tilde t$ limit  [GeV] & figure\\
\hline
$\tilde t_L +\tilde b_L$ & $\tilde H$ & $\sim 250$ & \ref{fig:StopHiggsinoLimit}    \\
$\tilde t_R$ & $\tilde H$ & $\sim 180$&  \ref{fig:StopHiggsinoLimit}  \\
$\tilde t_L + \tilde b_L$ & $\tilde B$ & $\sim 250-350$ & \ref{fig:StopBinoLimit} \\
\hline
\end{tabular}
\end{center}
\caption{\label{tab:StopResults}
A summary of the limits we found on direct stop and left-handed sbottom production with higgsino and bino LSPs.  The full limits are shown in the listed figures and the parameter spaces are described in the text of section~\ref{subsec:stop}.
}
\end{table}

\begin{table}[h!]
\begin{center}
\begin{tabular}{|c||c|c|c|}
\hline
scenario & $\tilde g$ limit  [GeV] & $\tilde t$ limit  [GeV] & figure \\
\hline
$\tilde H$ - LSP & $\sim 650-700$ & $\sim 280$ & \ref{fig:StopGluinoLimit}\\
$\tilde B$ - LSP & $\sim 700$ & $\sim 270$	& \ref{fig:StopGluinoLimit}		\\
somewhat squashed & $\sim 600-700$ & $-$ 	& \ref{fig:OtherStopGluinoLimit}		\\
split $\tilde t$ & $\sim 550-650$ & $-$	&	\ref{fig:OtherStopGluinoLimit}	 \\
flavor degen. & 1200 (fixed) & $600-900$ 	&	 \ref{fig:Unify}\\
gaugino unify & $\sim750-800$ & $\sim 260$ 	&	 \ref{fig:Unify}	\\
\hline
\end{tabular}
\end{center}
\caption{\label{tab:GluinoResults}
A summary of limits that we found in scenarios with gluinos.  The full limits are shown in the listed figures and the parameter spaces are described in the text of sections~\ref{subsec:gluino} and~\ref{sec:interpretation}.
}
\end{table}

On one hand we find that the current searches already started probing the direct production of third generation squarks, mostly in the $b+\chi$ decay channel. On the other hand, we find similar bounds on gluinos decaying through third generation squarks as those found by the experimental collaborations, but with the striking feature that tailored searches for gluinos decaying into heavy flavor squarks are currently not providing the most stringent bounds.

We do not attempt to make any future projections for the mass reach for stops, bottoms, higgsinos and the gluino for 5 and 10 $fb^{-1}$ of LHC data. The main reason is that the largest gain in reach will be likely come from new analyses designed and optimized for the parameter space regions where the current analyses are less powerful. Designing such analyses is beyond the scope of this work, and it requires a detailed study of the backgrounds, some of which, such as fakes, cannot be reliably estimated in a theoretical paper. Moreover, even the pure extrapolation of the reach of the current searches is plagued by intrinsic difficulties, not unrelated to those relevant for designing new analyses, which are discussed in Appendix~\ref{app-projections}.

We conclude by observing that the experimental program of searches for supersymmetry is crossing an important milestone.  The current searches are passing the naturalness threshold for stops and gluinos, and this means that the most favored parameter space of supersymmetry is just ahead of us.  If supersymmetry exists at the weak scale in a natural form, then discovery should be imminent.  On the other hand, if the LHC experiments fail to discover supersymmetry in the natural parameter space then, as the fine-tuning is increased, exotic manifestations of supersymmetry that are less constrained, such as hadronic $R$-parity violation~\cite{Barbier:2004ez} or stealth SUSY~\cite{stealth}, will become increasingly more interesting alternatives, both theoretically and experimentally.  The next frontier may be heavy-flavor-themed naturalness, or exotic searches.  Either way, the LHC will cover very exciting ground over the coming years.

\emph{Note added}: While this work was being completed, the authors of~\cite{thirdgenfriends1,thirdgenfriends2,thirdgenfriends3} informed us about related but distinct collider studies involving third generation squarks.
 
\begin{acknowledgments}
We acknowledge P.~Schuster and N.~Toro for participation at an early stage of this work.
We thank M.~R.~d'Alfonso, J.-F.~Arguin, S.~Caron, B.~Heinemann, A.~Hoecker, S.~A.~Koay, M.~d'Onofrio, S.~Padhi, M.~Pierini, P.~Pralavorio, G.~Redlinger, C.~Rogan, R.~Rossin, M.~Spiropulu, and I.~Vivarelli  for many suggestions and patiently answering our questions about the ATLAS and CMS searches.  We also thank N.~Arkani-Hamed, R.~Barbieri, C.~Cheung, S.~Dimopoulos, G.~F.~Giudice, L.~J.~Hall, I.~Low, M.~Perelstein, G.~Weiglein and N.~Weiner for discussions. M.P. and A.W. thank E.~Gianolio  for computing support.  J.T.R. thanks the Institute for Advanced Study for kindly providing access to the Aurora Cluster.  The work of M.P. was supported in part by the US Department of Energy under Contract DE-AC02-05CH11231. J.T.R. is supported by a fellowship from the Miller Institute for Basic Research in Science.  M.P. and J.T.R. would like to thank the Aspen Center for Physics where part of this work was conducted. The work of A.W. was supported in part by the German Science Foundation (DFG) under the Collaborative Research Center (SFB) 676.

\end{acknowledgments}

\appendix

\section{Validation of the analyses implementations}\label{app:validation}

In order to check whether our PGS/Atom implementations are giving results in reasonable agreement with those obtained by the experimental collaborations, for each analyses we validated them by comparing with the publicly available data. There are two kind of plots that one can compare the results to: kinematic distributions and exclusion limits.

\begin{figure}[h!]
\label{fig:calibrate}
\begin{center} \includegraphics[width=1.1 \textwidth]{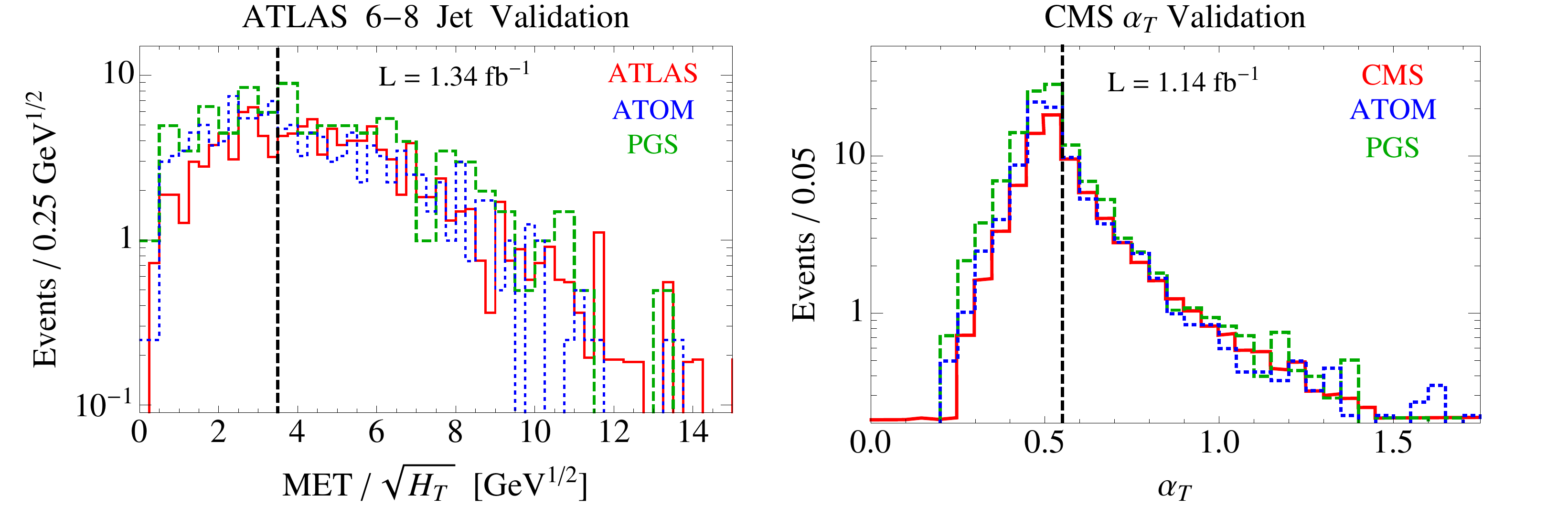} \end{center}
\caption{Validation of kinematic plots for ATOM and PGS.  The {\it left} plot shows the missing energy significance in the ATLAS 6-8 jets plus missing energy search, for the MSUGRA benchmark point with $m_0 = 1220$~GeV and $m_{1/2} =  180$~GeV, $\tan \beta = 10$, and $A_0=0$~GeV.  The {\it right} plot shows the distribution of $\alpha_T$ for the CMS search using this variable and the MSUGRA benchmark point LM6.  In both plots, the signal region is to the right of the vertical black dashed line, and we find good agreement between the experimental simulations, and ATOM and PGS.}
\end{figure}

In the first case the event distribution for a particular observable is plotted for a specific signal model and a specific point in parameter space. Comparing against such a histogram is very useful to detect kinematic distortions induced by our approximations (from the shape of the distribution) and to compare precisely the signal acceptances and efficiencies, $\epsilon \times A$ (from the histogram normalization). Examples of such comparisons are shown in Fig.~\ref{fig:calibrate} for two different cases: the $\met$ significance for the ATLAS 6-8 jets+$\met$ search and the $\alpha_{T}$ distribution for the CMS search. As one can see both of our pipelines reproduce reasonably well the kinematic distributions and  acceptances of hadronic SUSY searches without the need of further adjustments, which is important since many of our limits depend on jets+$\met$ analyses.

\begin{figure}[h!]
\label{fig:cali-limits}
\begin{tabular}{S S}
\vspace{.5cm}\includegraphics[width=.65 \textwidth]{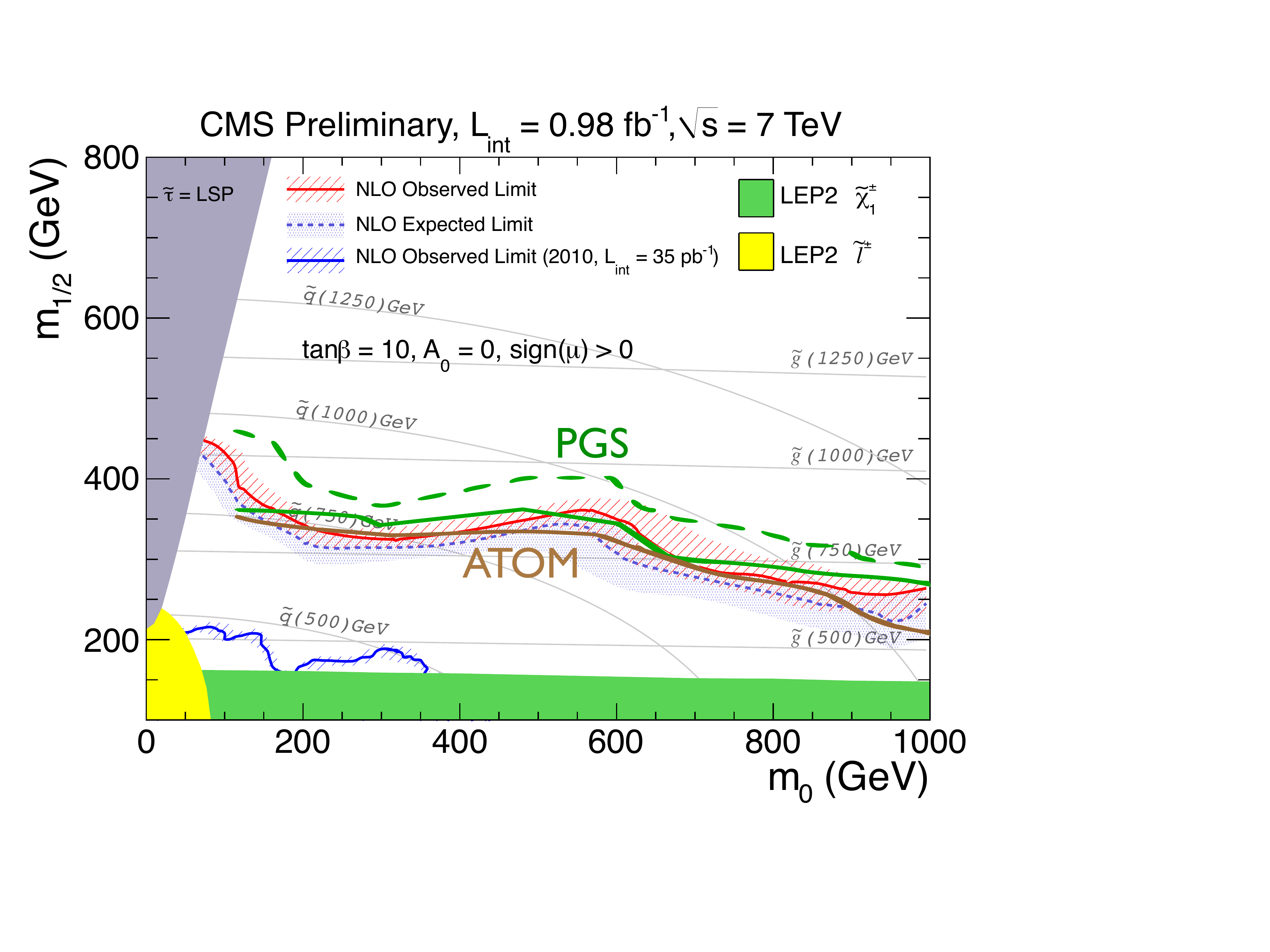} &  \includegraphics[width=0.55 \textwidth]{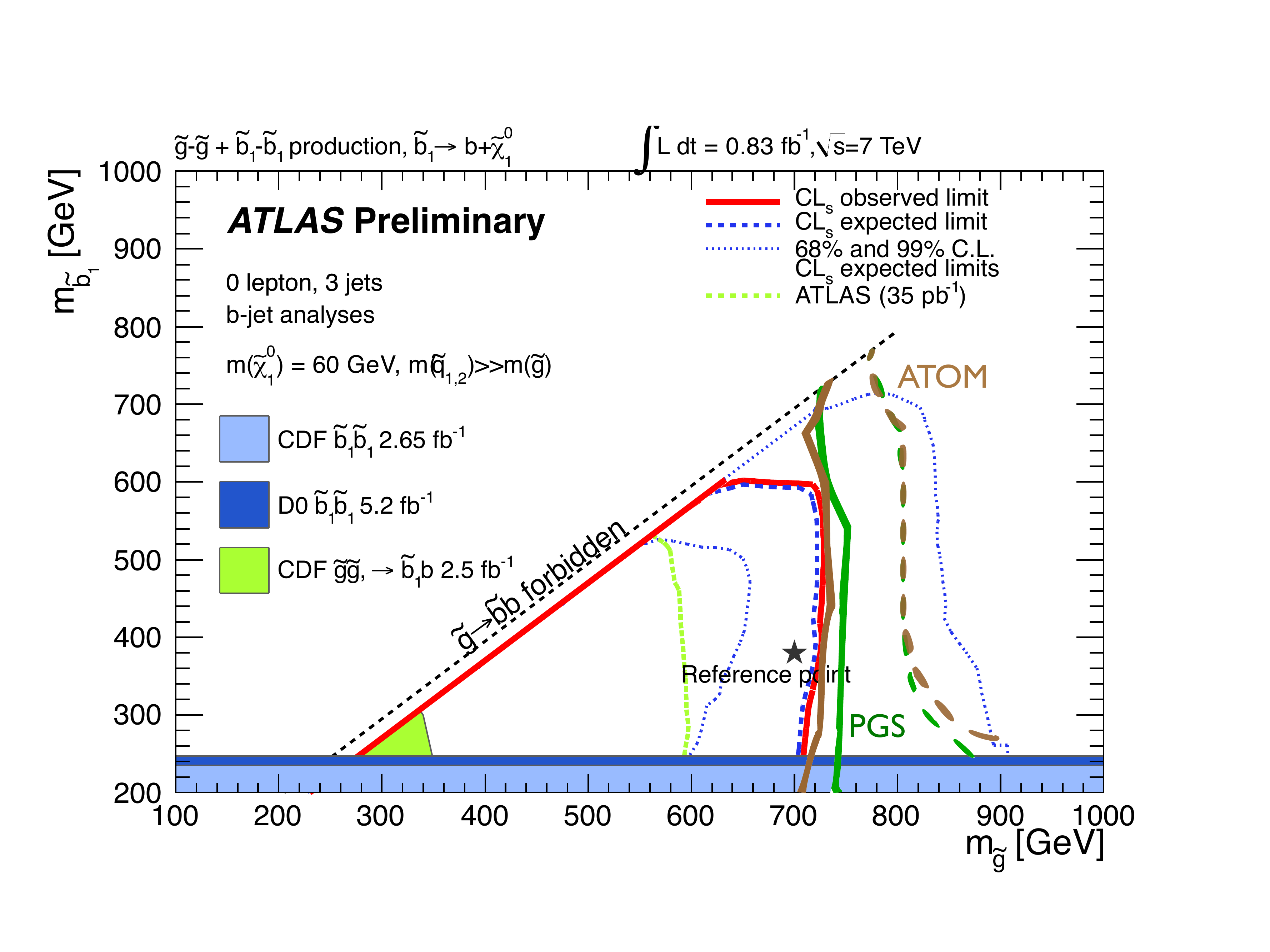} 
\end{tabular}
\caption{Validation of exclusion limit plots for ATOM and PGS.  The {\it left} plot shows the CMSSM limit for the Same-Sign dilepton search by CMS, and superimposed the PGS (green) and ATOM (brown) curves. The dashed curve represent the PGS prediction before correcting for the difference in lepton identification efficiencies between the code (90\%) and the CMS analysis (roughly 70\%), while the solid line correspond to the final result.
The {\it right} plot shows instead the exclusion limit for the gluino-sbottom-neutralino simplified model presented in the b-jets+$0\ell$+$\met$ ATLAS analyses. PGS (ATOM) curves are shown in green (brown), where the dashed line is the limit before the factor of 2 correction on the event yield due to the systematic uncertainties on the signal, and the solid line is the final result.}
\end{figure}

One drawback is that the comparison is for a specific point in parameter space, therefore one cannot detect potential problems in different kinematical regions. A different cross-check is instead provided by exclusion plots, such as the simplified models or the classic limit in the CMSSM plane. In many cases these are the only plots one can compare to. Here the curves represent mass limits, which are often easier to match given that the steeply falling cross-section tend to reduce the effects of a discrepancy in $\epsilon \times A$. On the other hand such comparisons have the ability to check the agreement of our implementations in different kinematical regions at once.  However other sources of disagreement may appear and they render the process of debugging discrepancies considerably harder. A typical example is the effect of including systematic uncertainties on the signal in order to produce the limit, which typically introduce an intrinsic uncertainty in the comparison due to lack of information. In Fig.~\ref{fig:cali-limits} one can see the results for two of such comparisons, the mSUGRA limit for the Same-Sign dilepton  CMS search and the ATLAS  bjets+0leptons+$\met$ analysis. In particular the latter analysis also shows the stronger level of discrepancy (a factor of 2 in $\epsilon \times A$) among all our comparisons, most likely due to systematics on the signal we did not include. However we did check, by using a crude estimate of their size from~\cite{ATLAS-CONF-2011-098}, that the $CL_{s}$ limits on event yields may vary by a factor of two. Therefore we decided to apply this correction factor everywhere in our study. Fig.~\ref{fig:cali-limits} shows the effects of this rescaling.

\section{Brief description of ``ATOM''}\label{app:atom}

ATOM (``Automatic Test Of Models'') is the tentative name of a tool currently developed by some of the authors and it is intended to be released in the future for the free use to the community. The purpose of such tool is to provide, by running locally on the user's computer, a relatively fast approximate (although often ``good enough'') answer to the question whether a specific model is excluded or not by a set of experimental searches. It does not aim to provide the full correct answer, which can be provided only by a real study by the experimental collaborations or by more powerful tools like RECAST~\cite{recast} currently under development. A detailed description of the package will be given elsewhere~\cite{atom}, here we will just highlight the main features. The tool accepts particle events as a definition for the model currently being tested. The event processing is performed by the Rivet package~\cite{rivet} upon which ATOM is built. An advantage of Rivet is that a large number of analyses can be performed simultaneously without a significant extra cost in CPU time. As in the base version of Rivet, ATOM processes the input events through the cuts of the implemented analyses and  populates the various histograms present in the various experimental papers. 

For the analyses we have coded, we included also the various plots corresponding to the control regions used by the analyses to determine the backgrounds. This is important in order to check whether a new physics signal may substantially leak into a control region for a search and be ``subtracted away'', especially if the latter has not been specifically designed for that particular model. Differently than the base version of Rivet, ATOM automatically saves the information about signal efficiencies at various stages of the analyses, both for the total signal events and for each individual sub-process. Moreover, for each cut, it automatically computes the sensitivity of the signal efficiency to the precise value of that cut (defined as the logarithmic derivative of the efficiency with respect to the cut position). We use this feature to detect regions where the cuts are applied on steeply falling signal distributions, leading to large uncertainties in the final efficiency as, \eg, in Fig.~\ref{fig:StopGluinoLimit}.

All this additional information in addition to the Rivet histograms is parsed by ATOM to flag potential problems for the results of the analyses with the signal events at hand. The final efficiencies are then used in the statistics module to extract the exclusion limits.

The events are processed by default at truth level as in Rivet. Jets are clustered with FastJet~\cite{fastjet}. We perform lepton isolation at particle level according to the parameters specified in the experimental papers and we reconstruct b-jets by determining whether the particles clustered in a jet have a b-quark ancestor and then applying a tagging efficiency as specified by the searches. 

We have implemented in ATOM also the possibility to use parameterised efficiency specified as 2D histograms in $p_{T}$ and $\eta$ for all the various objects, as well as the possibility of including smearing. However we do not use them in the present study and we limit ourselves to apply the reconstruction efficiency for leptons as a constant correction factors whenever specified in the papers.

\section{Projections for the current analyses}\label{app-projections}

\begin{figure}[h!]
\begin{center} \includegraphics[width=0.5 \textwidth]{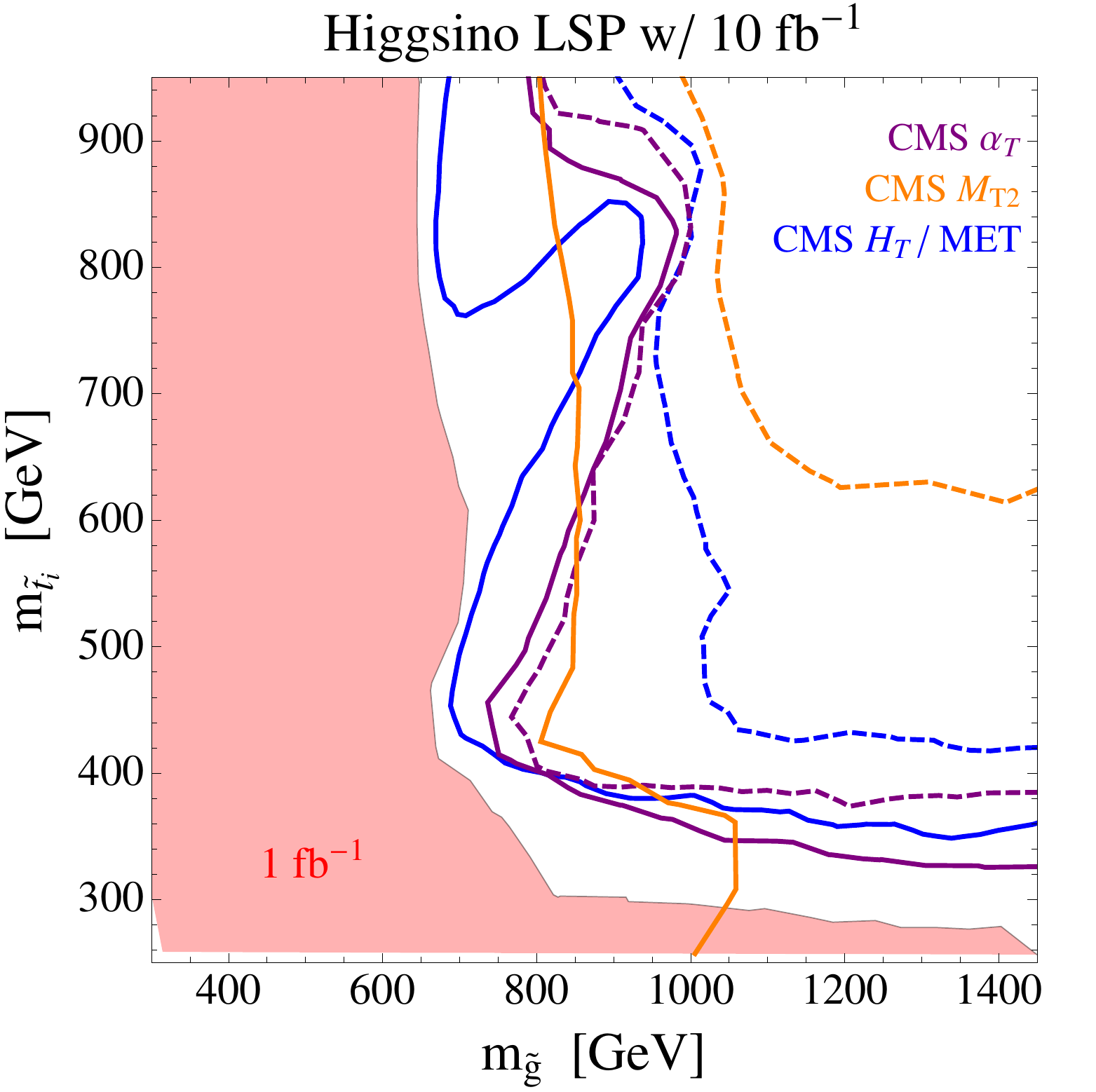} \end{center}
\caption{\label{fig:projections}
Possible range for the projections of the current analyses to $10fb^{-1}$ of LHC data in the case of gluinos and stops decaying to higgsino LSP. The solid lines correspond to the conservative assumption of rescaling the statistical errors with the luminosity and keeping the relative systematic as constant, while the dashed lines correspond to the extremely optimistic case of perfect knowledge of the backgrounds.}
\end{figure}

Here we discuss the (im)possibility of extrapolating to higher luminosities the reach of the current analyses, given the limitations of our ``theorist'' analysis. The most naive (and conservative) extrapolation would be to scale the statistical errors with the increased luminosity and keep the relative systematic error as constant. However one notices immediately that in most of the analyses the systematic errors on the backgrounds are of the same order as the statistical ones. Therefore even with a large increase in luminosity the limit on the cross-section would improve only by a factor of $\sim \sqrt 2$, which corresponds to a limited increase in the mass reach. This is unlikely to be the case, the reason being that in most of the cases the systematic errors have been currently reduced to be a subdominant component of the error budget, even if there may still be the possibility of further improvements. The correct procedure would be to study in detail the systematic error budget and estimate for each of them what would be the improvement in the future, a task clearly beyond the scope of this paper. On the other extreme, one could try to understand what would be the upper limit on the improvement by (unrealistically) assuming a perfect knowledge of the background and include only the Poissonian error in computing the limits. Obviously the correct answer lies in between these two extrema, but as one can see from Fig.~\ref{fig:projections} the mass range spanned by these two limits is extremely large, rendering useless any projections done with our means. There is another reason for avoiding any attempt for giving projections: in many cases the backgrounds in the signal regions are determined by control regions and therefore are sensitive to statistical fluctuations there in the current dataset. This is the case, \eg, for the CMS $M_{T_{2}}$ analyses where, as stated in~\cite{CMS-PAS-SUS-11-005} a downward fluctuation in the last bin of the control region, have determined a lower background estimate in the signal region. Extrapolating the projections to $10fb^{-1}$ would yield very powerful constraints as shown in Fig.~\ref{fig:projections}, that would be completely overestimated if indeed the low background is due to a statistical fluctuation.

\end{document}